\newif\ifpreprint

\preprintfalse % Enable for double column reprint

\ifpreprint
\documentclass[journal=jctcce,manuscript=letter]{achemso}
\else
\documentclass[journal=jctcce,manuscript=letter,layout=twocolumn]{achemso}
\fi

\usepackage[T1]{fontenc} % Use modern font encodings

\usepackage{amsmath}
\usepackage{newtxtext,newtxmath}

\usepackage{graphicx}
\usepackage{dcolumn}
\usepackage{braket}
\usepackage{multirow}
\usepackage{threeparttable}
\usepackage{xspace}
\usepackage{verbatim}
\usepackage[version=4]{mhchem} % Formula subscripts using \ce{}
\usepackage{comment}
\usepackage{color,soul}

\usepackage{mathtools}
\usepackage[dvipsnames]{xcolor}
\usepackage{xspace}
\usepackage{ifthen}

\usepackage{qcircuit}

\usepackage{graphicx,longtable,dcolumn,mhchem}
\usepackage{rotating,color}
\usepackage{lscape}
\usepackage{amsmath}
\usepackage{dsfont}
\usepackage{soul}
\usepackage{physics}
\newcolumntype{d}{D{.}{.}{-1}}

\newcommand{\AVDZ}{\emph{aug}-cc-pVDZ}
\newcommand{\AVTZ}{\emph{aug}-cc-pVTZ}
\newcommand{\AVQZ}{\emph{aug}-cc-pVQZ}
\newcommand{\DZ}{cc-pVDZ}
\newcommand{\TZ}{cc-pVTZ}
\newcommand{\QZ}{cc-pVQZ}

\usepackage[colorlinks = true,
            linkcolor = blue,
            urlcolor  = black,
            citecolor = blue,
            anchorcolor = black]{hyperref}

\definecolor{goodorange}{RGB}{225,125,0}
\definecolor{goodgreen}{RGB}{5,130,5}
\definecolor{goodred}{RGB}{220,50,25}
\definecolor{goodblue}{RGB}{30,144,255}
\renewcommand{\hl}[1]{\textcolor{black}{#1}}

\newcommand{\note}[2]{
\ifthenelse{\equal{#1}{F}}{
\colorbox{goodorange}{\textcolor{white}{\footnotesize \fontfamily{phv}\selectfont #1}}
    \textcolor{goodorange}{{\footnotesize \fontfamily{phv}\selectfont #2}}\xspace
}{}
\ifthenelse{\equal{#1}{R}}{
\colorbox{goodred}{\textcolor{white}{\footnotesize \fontfamily{phv}\selectfont #1}}
    \textcolor{goodred}{{\footnotesize \fontfamily{phv}\selectfont #2}}\xspace
}{}
\ifthenelse{\equal{#1}{N}}{
\colorbox{goodgreen}{\textcolor{white}{\footnotesize \fontfamily{phv}\selectfont #1}}
    \textcolor{goodgreen}{{\footnotesize \fontfamily{phv}\selectfont #2}}\xspace
}{}
\ifthenelse{\equal{#1}{M}}{
\colorbox{goodblue}{\textcolor{white}{\footnotesize \fontfamily{phv}\selectfont #1}}
    \textcolor{goodblue}{{\footnotesize \fontfamily{phv}\selectfont #2}}\xspace
}{}
}

\usepackage{titlesec}

\usepackage[fontsize=11pt]{scrextend}
\titleformat{\section}
{\normalfont\sffamily\bfseries\color{Blue}}
{\thesection.}{0.25em}{\uppercase}

\titleformat{\subsection}
{\normalfont\sffamily\bfseries}
{\thesubsection}{0.25em}{}

\titleformat{\subsubsection}
{\normalfont\sffamily}
{\thesubsubsection}{0.25em}{}

\titleformat{\suppinfo}
{\normalfont\sffamily\bfseries}
{\thesubsection}{0.25em}{}

\titlespacing*{\section}{0pt}{0.5\baselineskip}{0.01\baselineskip}
\titlespacing*{\subsection}{0pt}{0.125\baselineskip}{0.01\baselineskip}
\titlespacing*{\subsubsection}{0pt}{0.125\baselineskip}{0.01\baselineskip}

\author{Pierre-Fran\c{c}ois Loos}
	\affiliation[LCPQ, Toulouse]{Laboratoire de Chimie et Physique Quantiques, Universit\'e de Toulouse, CNRS, UPS, France}
	\email{loos@irsamc.ups-tlse.fr}
\author{Massimiliano Comin}
	\affiliation[NEEL, Grenoble]{Univ. Grenoble Alpes, CNRS, Inst NEEL, F-38042 Grenoble, France}
\author{Xavier Blase}
	\affiliation[NEEL, Grenoble]{Univ. Grenoble Alpes, CNRS, Inst NEEL, F-38042 Grenoble, France}
	\email{xavier.blase@neel.cnrs.fr}
\author{Denis Jacquemin}
	\email{Denis.Jacquemin@univ-nantes.fr}
	\affiliation[CEISAM, Nantes]{Universit\'e de Nantes, CNRS,  CEISAM UMR 6230, F-44000 Nantes, France}
		
\let\oldmaketitle\maketitle
\let\maketitle\relax
     \title{Reference Energies for Intramolecular Charge-Transfer Excitations}
\date{\today}

\begin{tocentry}
	\centering
	\includegraphics[width=.55\textwidth]{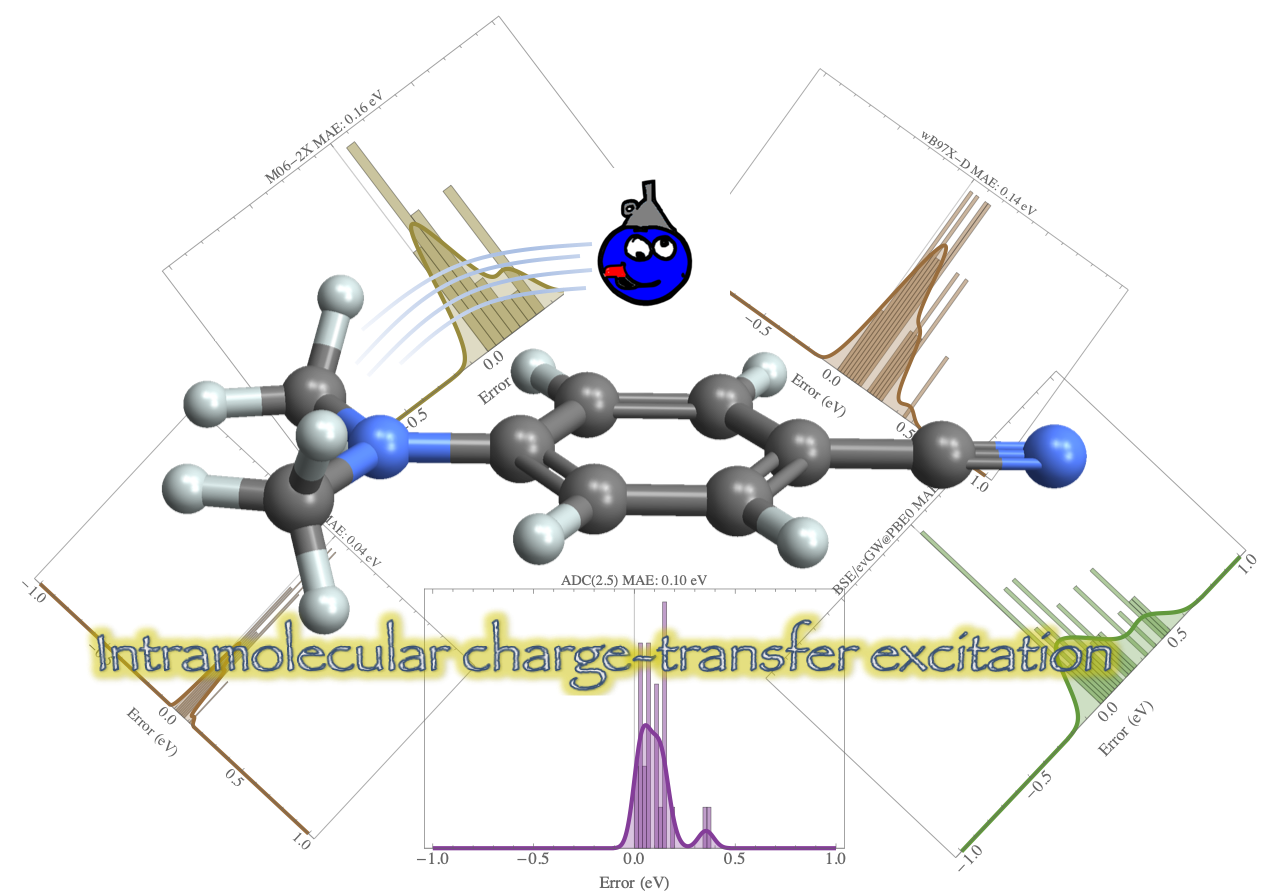}
\end{tocentry}

\begin{document}

\ifpreprint
\else
\twocolumn[
\begin{@twocolumnfalse}
\fi
\oldmaketitle

%%%%%%%%%%%%%%%%
%%% ABSTRACT %%%
%%%%%%%%%%%%%%%%
\begin{abstract}
In the aim of completing our previous efforts devoted to local and Rydberg transitions in organic compounds, we provide a series of highly-accurate vertical transition energies for intramolecular charge-transfer transitions 
occurring in ($\pi$-conjugated) molecular compounds. To this end we apply a composite protocol consisting of linear-response CCSDT excitation energies determined with Dunning's double-$\zeta$ basis set corrected by 
CC3/CCSDT-3 energies obtained with the corresponding triple-$\zeta$ basis. Further basis set corrections (up to \emph{aug}-cc-pVQZ) are obtained at the \hl{CCSD and} CC2 level. We report 30 transitions obtained in 
17 compounds (ABN, aniline, azulene, benzonitrile, benzothiadiazole, DMABN, dimethylaniline, dipeptide, $\beta$-dipeptide, hydrogen chloride, nitroaniline, nitrobenzene, nitrodimethylaniline, nitropyridine N-oxide, 
N-phenylpyrrole, phthalazine,  and quinoxaline).   These reference values are then used to benchmark a series of wave function (CIS(D), \hl{SOPPA, RPA(D),} EOM-MP2, CC2, CCSD, CCSD(T)(a)*, CCSDR(3), CCSDT-3, 
CC3, ADC(2), ADC(3), and ADC(2.5)), the Green's function-based Bethe-Salpeter equation (BSE) formalism performed on top of the partially self-consistent ev$GW$ scheme considering two different starting points 
(BSE/ev$GW$@HF and BSE/ev$GW$@PBE0), and TD-DFT combined with several exchange-correlation functionals  (B3LYP, PBE0, M06-2X, CAM-B3LYP, LC-$\omega$HPBE, $\omega$B97X, $\omega$B97X-D, and M11). 
It turns out that \hl{the CC methods including triples, namely,} CCSD(T)(a)*, CCSDR(3), CCSDT-3, \hl{and} CC3 provide rather small average deviations ($\leq 0.10$ eV), CC3 emerging as the only chemically accurate approach. 
\hl{ADC(2.5) also performs nicely with a mean absolute error of 0.11 eV for a} $\mathcal{O}(N^6)$ \hl{formal scaling, whereas} CC2 and  BSE/ev$GW$@PBE0 also deliver  \hl{very} satisfying results given their respective $\mathcal{O}(N^5)$ and 
$\mathcal{O}(N^4)$ computational scalings. In the TD-DFT context, the best performing functional is $\omega$B97X-D, closely followed by CAM-B3LYP 
and M06-2X, all providing mean absolute errors around $0.15$ eV relative to the theoretical best estimates.  
\end{abstract}

\ifpreprint
\else
\end{@twocolumnfalse}
]
\fi

\ifpreprint
\else
\small
\fi

\noindent

%%%%%%%%%%%%%%%%%%%%%%%%%%%%%%%%%%%%%%
\section{Charge-transfer excitations}
\label{sec:CTex}
%%%%%%%%%%%%%%%%%%%%%%%%%%%%%%%%%%%%%%
Charge-transfer (CT) transitions are key to the working principle of many practical applications of photoactive molecules (OLEDs, photovoltaics, photosynthesis, ion probes, etc). For this reason they have been widely studied, and they are generally regarded as a specific class of excitations, 
fundamentally different from their more local valence and Rydberg counterparts. While there is no formal definition of CT, chemists generally consider that, in a CT transition, the excitation process transfers a significant fraction of electron density from one molecular fragment, the donor $\text{D}$, 
to another fragment, the acceptor $\text{A}$. These two fragments can be part of the same molecule (intramolecular CT) or belong to two distinct molecules (intermolecular CT). The CT excitation induces a significant charge shift in going from the ground state (GS) to the excited state (ES), the latter being typically (much) more polar than the former. The reverse situation in which the dipole strongly decreases upon excitation can also be observed 
(e.g., in betaine 30 \cite{Cer14}).  In other words, CT transitions are characterized by a large change in dipole moment as well as a small overlap between the starting and final molecular orbitals (MOs), or electron densities, involved in the transition. 

From a more theoretical point of view, considering an overall neutral system, one can show that a CT excitation energy behaves, for large enough separation $R$ between the donor and acceptor (the so-called Mulliken limit \cite{Mul52}) as \cite{Dre05,Mai17}
\begin{equation}\label{eq:CT}
	\Delta E_\text{CT} = \text{IP}^\text{D} - \text{EA}^\text{A} - 1/R,
\end{equation}
where $\text{IP}^\text{D}$ is the first ionization potential of the donor, $\text{EA}^\text{A}$ is the electron affinity of the acceptor, and $-1/R$ is the electrostatic interaction between the excited electron located on the acceptor fragment and the hole left behind located on the donor fragment.
Due to the wrong asymptotic behavior of the kernel associated with (semi-)local exchange-correlation functionals (XCFs), it was quickly recognized that capturing the correct $-1/R$ asymptotic behavior of Eq.~\eqref{eq:CT} is particularly challenging for time-dependent density-functional theory 
(TD-DFT). \cite{Toz03,Dre04} Furthermore, the energy difference between the donor ionization potential ($\text{IP}^\text{D}$) and acceptor electron affinity ($\text{EA}^\text{A}$) tends to be too small when using Kohn-Sham (KS) orbital energies obtained with semi-local functionals due to the lack of derivative discontinuity into the XCF upon electron addition or removal. \cite{Per83}  As a result,  in its traditional adiabatic formulation, TD-DFT tends to drastically underestimate CT transition energies when combined with local-density approximation (LDA) or generalized-gradient approximation (GGA) XCFs. Some improvements 
are observed for global hybrid functionals (such as B3LYP \cite{Bec88,Lee88,Bec93} and PBE0 \cite{Ada99,Erz99}) that combine a uniform fraction of Hartree-Fock (HF) exchange with a (semi-)local XCF.  However, unless 100\% of exact exchange is included, a systematic underestimation 
of CT excitation energies remains due to the relative short-sightedness of global hybrids. Historically, this limitation strongly motivated the development of range-separated hybrids (RSHs) \cite{Sav96,IIk01,Yan04,Vyd06,Cha08} and their optimally-tuned versions \cite{Ste09,Kro12} which provide a 
more subtle blend by switching gradually, as a function of the interelectronic distance, from short-range (semi-)local exchange to long-range HF exchange. In such a way, one can combine the best of both worlds by benefiting from the short-range dynamical correlation effects given by  DFT as 
well as key error cancellation between exchange and correlation, while using 100\% HF exchange at long range in order to entirely take into account the interaction between the electron and the hole, hence capturing the correct $-1/R$ asymptotic behavior. RSHs are particularly effective at describing 
CT transitions, but often at the cost of a (slight) overestimation of the transition energies of the corresponding local excitations.\cite{Lau13}  

As an alternative to TD-DFT, one can use  the Bethe-Salpeter equation (BSE) formalism~\cite{Sal51,Han79,Roh98,Alb98b,Ben98,Hor99,Bla18,Bla20} starting with $GW$ quasiparticle energies~\cite{Hed65,Str80,Hyb86,God88,Oni02,Pin13b,Gol19} (BSE/$GW$) as specific formulations of Green's 
function many-body perturbation theory, By construction, this scheme explicitly includes terms describing the non-local electron-hole interactions,  together with an accurate description at the $GW$ level of the ionization potentials and electron affinities, allowing to ``naturally'' deliver accurate CT energies 
for a computational cost comparable to TD-DFT.  \cite{Bla11,Bau12b,Duc12,Loo20c,Bla20} The description of the non-local electron-hole interaction thanks to the screened Coulomb potential $W$ allows to consider CT excitations in situations differing from the ideal long-range CT through vacuum, 
a property central to the study of intramolecular CT or CT in effective dielectric media such as organic semiconductors. \cite{Pus02,Tia03,Cud13,Bau14,Li17,Duc18}  Of course, one can also turn towards wavefunction approaches, and both the second-order algebraic diagrammatic construction 
[ADC(2)] \cite{Tro97,Dre15} and approximate second-order coupled-cluster method (CC2)  \cite{Chr95,Hat00} methods are generally regarded as well-suited for accurately describing CT phenomena.

%%%%%%%%%%%%%%%%%%%%%%%%%%%%%%%%%%
\section{Charge-transfer metrics}
\label{sec:CTmet}
%%%%%%%%%%%%%%%%%%%%%%%%%%%%%%%%%%
How does one pinpoint a CT transition? Experimentally, the identification of CT transitions is typically achieved by investigating the absorption spectrum: a strong CT  induces a large increase of dipole moment when 
going from the GS to the ES, which in turn, translates into a broad and structureless absorption band undergoing significant redshifts when the polarity of the solvent increases (the so-called positive solvatochromism). 
Theory obviously delivers a complementary view for unveiling CT states. A decade ago, such task was often performed by investigating the topology of the MOs involved in the transition and/or the changes 
of partial atomic charges following the electronic excitation. Such analyses were certainly successful, but they obviously lacked systematic character. Hence, more quantifiable metrics have been recently developed. 

The first we are aware of is the so-called $\Lambda$ parameter defined by Tozer in 2008. \cite{Pea08} $\Lambda$ measures the overlap between the occupied and virtual orbitals involved in a specific transition, and was originally 
applied by the Tozer group to demonstrate the superiority of RSHs for CT and Rydberg transitions. \cite{Pea08} In 2011, Le Bahers, Adamo, and Ciofini came up with the $d^{\text{CT}}$ metric, \cite{Leb11c} 
which measures the distance between the barycenters of density gain and depletion upon excitation; this model is thus particularly well-acquainted to density-based approaches. \cite{Ada15,Hue20}  Following these two seminal works, 
many other strategies have been proposed to quantify CTs, such as Guido's $\Delta r$, \cite{Gui13b} which measures the electron-hole distance thanks to an analysis of the charge centroids of the orbitals involved in 
the excitation, Etienne's $\phi_s$ which is based on the detachment/attachment matrices, \cite{Eti14} and the more general approaches developed by Dreuw's group, \cite{Pla14,Pla14b} which allow analyses 
not only at the TD-DFT level but also with more advanced wavefunction theories such as ADC(2). Several of these metrics have been implemented in well-known quantum chemistry codes and clearly enjoy a strong 
popularity in the community.  In this framework, we specifically highlight the purpose-designed TheoDORE package \cite{Pla20} encompassing many models for investigating ES topologies.  

Although these various metrics do not provide a definite answer to the \textit{``what is a CT transition''} question, and potentially deliver distinct answers depending on the nature of the underlying (density or wavefunction) description, 
they nevertheless offer a large panel of options for quantifying the CT strength.

%%%%%%%%%%%%%%%%%%%%%%%%%%%%
\section{Literature survey}
\label{sec:CTlit}
%%%%%%%%%%%%%%%%%%%%%%%%%%%%

To evaluate the performances of specific methods for CT transitions, various sets of reference values have been proposed over the years. Let us describe a selection of some relevant datasets. 

In their seminal TD-DFT work, \cite{Pea08}  Peach and coworkers gathered a group of 14 intramolecular CT transitions obtained in three model peptides, N-phenylpyrrole (PP), dimethylamino-benzonitrile (DMABN) and hydrogen chloride (HCl). The reference 
values were taken from a previous CASPT2 \cite{And90,And92} work for the peptides, \cite{Ser98b} extracted from experiment for DMABN, and determined at the CC2 level for both PP and HCl. The same reference values were
used in the following years to assess various DFT approaches. \cite{Ngu11,Mar11,Hed13} However, in 2012, the Tozer group used EOM-CCSD to define new benchmark transition energies for the smallest peptide as well 
as for both planar and twisted DMABN and PP,  in a work encompassing 9 reference CT  energies.\cite{Pea12} The same year, Dev, Agrawal and English compiled a set of 16 CT transitions in large conjugated dyes, \cite{Dev12} and 
they exclusively employed experimental data as reference.   In 2015, He{\ss}elmann considered the 10 CASPT2 values obtained for peptides \cite{Ser98b} and the 2 EOM-CCSD data determined for PP \cite{Pea12} as benchmarks for evaluating the 
performance of non-standard TD-DFT schemes. \cite{Hes15b} All the transitions of the original contribution of Peach and coworkers were re-evaluated in 2019 by Goerigk's group, 
\cite{Cas19} which proposed updated references obtained with CCSDR(3)/{\TZ} \cite{Chr96b} (or SCS-CC2) \cite{Hel08} often using a basis set extrapolation technique similar to the one applied here (\textit{vide infra}). This set was completed by three 
additional cases, namely \emph{para}-nitro-aniline (pNA) with a reference value obtained with CCSDR(3), the benzene-tetracyanoethylene (B-TCNE) intermolecular complex with a EOM-CCSD(T) \cite{Wat95} reference, and a 
large dye (so-called DBQ)  for which experiment was used as benchmark.\cite{Goe10a} These variations in reference values along the years for Tozer's original set clearly highlight the appetite of the electronic structure community 
for high-quality benchmark values, as well as the lack of indisputability for such data even for thoroughly-studied systems. 

We wish also to mention that the test set developed by Truhlar and Gagliardi \cite{Hoy16} contains three CT 
transitions for pNA [computed at the  $\gamma$-CR-EOM-CC(2,3)D level], \cite{Pie02b} DMABN (experiment) and B-TCNE (experiment). Two of us considered the same CT transitions to explore the performances of BSE/$GW$. 
\cite{Jac17b}  In 2018, Gui,  Holzer and  Klopper, \cite{Gui18b} used a set of seven CT states in pNA, DMABN, PP, HCl and B-TCNE in the similar context.
For the first five molecules, they proposed
basis set extrapolated CC3/{\AVTZ} results, therefore providing again new reference values for those popular systems. 

Other sets have been exclusively dedicated to intermolecular CT transitions, which can be viewed as conceptually 
simpler, as the electron ``jumps'' from one molecule to another during the CT excitation.  Such systems were used already 15 years ago by Truhlar's group, \cite{Zha05f,Zha08b} and have become very popular for benchmarking 
density- and wavefunction-based methods. \cite{Zha05f,Zha08b,Ste09,Aqu11,Sza13,Bla14,Gho15,Ott20,Koz20,Zul21} In 2009, Stein, Kronik and Baer used  13 experimental values measured in CT complexes constituted of 
an aromatic system interacting with TCNE to assess the performances of their optimally-tuned RSH functional. \cite{Ste09}  
The same systems were further studied at the BSE level in 2011. \cite{Bla11}
In 2020, Ottochian and coworkers followed a very similar strategy to benchmark various hybrid and double-hybrid functionals. 
\cite{Ott20}. In 2011, Aquino and coworkers employed ADC(2) as reference to benchmark various XCFs for CT occurring in stacked DNA bases. \cite{Aqu11} Similar stacked nucleobases were also studied  
by Szalay and coworkers in a 2013 work which reports EOM-CCSD(T) data, \cite{Sza13} in 2014 by Blancafort and Voityuk who obtained CASPT2 energies, \cite{Bla14} and in 2021 by the Matsika group which provided a large set of 
reference values obtained at the ADC(3)/cc-pVDZ level. \cite{Zul21}  

Again, the richness of reference values is obviously both an advantage and a drawback as it is objectively hard to know which work reports the most accurate transition energies.
Recently Kozma \emph{et al.}~tackled this question by defining 14 accurate intermolecular CT transitions obtained in molecular dimers (e.g., ammonia-fluorine, pyrrole-pyrazine, acetone-nitromethane, \ldots). \cite{Koz20} In this key work, 
the reference values are obtained at EOM-CCSDT  \cite{Kow01} or CCSDT-3 \cite{Wat96} levels (depending on the system size) with the {\DZ} basis set and several lower-order wavefunction approaches are benchmarked.  
Interestingly, this study reveals that for intermolecular CT transitions, CCSDT-3 is more accurate than CC3, \cite{Chr95b,Koc95} which is the opposite trends as  compared to local and Rydberg transitions. \cite{Ver21} To the very best of our 
knowledge, Ref.~\citenum{Koz20} stands today as the sole work providing reference CT values obtained at a very high level of theory (i.e., EOM-CCSDT).

%%% FIGURE 1 %%%
\begin{figure*}[htp]
\centering
 \includegraphics[width=.86\linewidth]{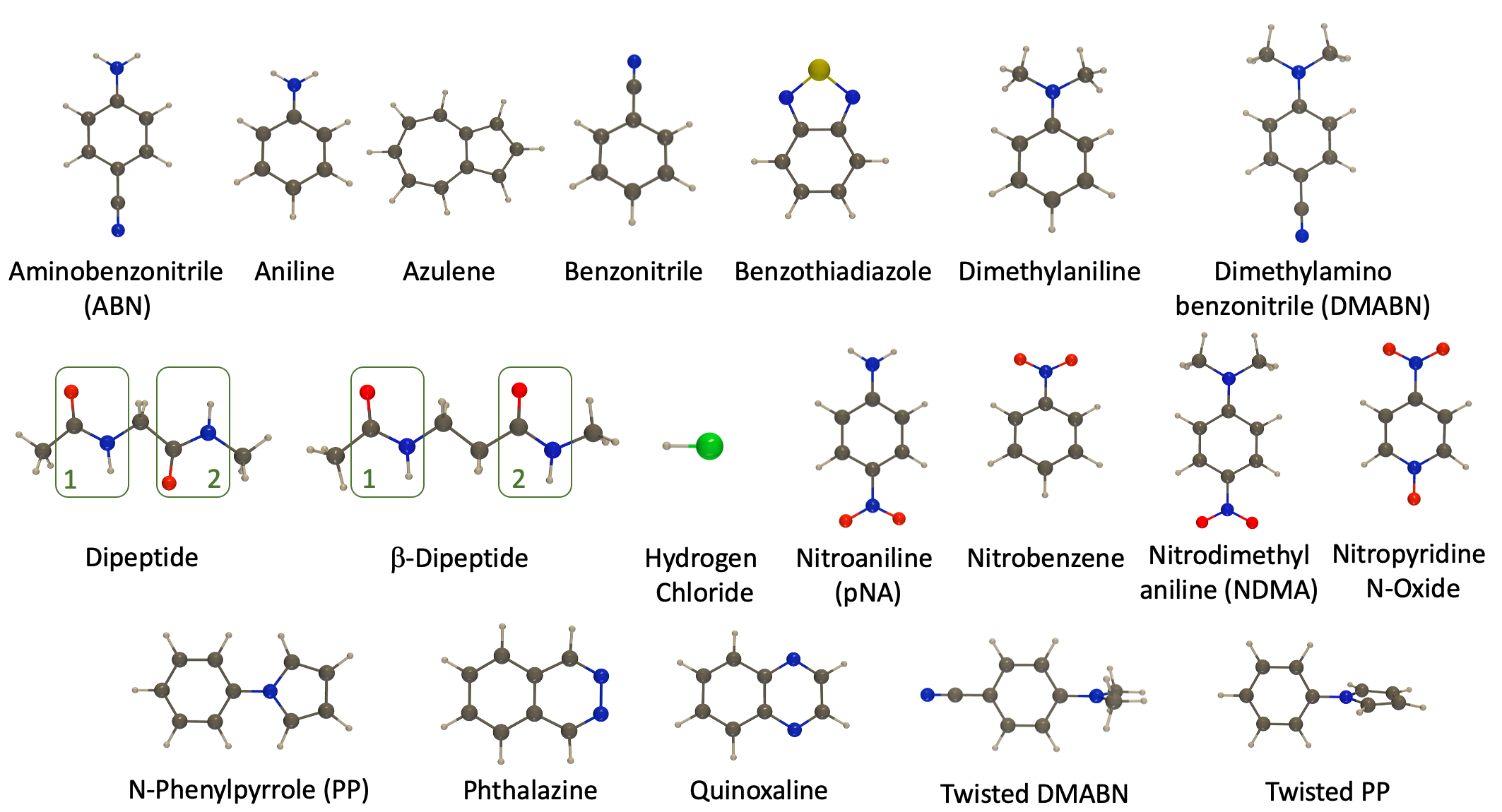}
  \caption{Representation of the investigated derivatives.}
   \label{Fig-1}
\end{figure*}
%%% %%% %%% %%%

Our goal here is to propose to the community a list of highly-accurate vertical transition energies for intramolecular CT excitations that can be used as reference to assess the \emph{pros} and \emph{cons} of lower-level models. 
For instance, there have been recent controversies in the literature regarding the relative accuracy of various double hybrids for CT transitions, \cite{Cas19,Ott20,Cas21,Mes21} whereas there are significant discrepancies (ca.~0.2 eV) 
between the recent CC-based theoretical best estimates (TBEs) obtained for pNA, DMABN, and PP by distinct groups, \cite{Gui18b,Cas19} and it is rather difficult to determine the actual origin (basis set, geometry, method, \ldots) of these differences.

We do hope that the present (rather large) set can help settling in these incertitudes. Obviously, some systems treated here have been taken from the sets described above, but we have both computed more accurate geometries (\emph{vide infra}) and clearly 
increased the level of theory employed to define the benchmark TBEs as compared to previous efforts devoted to intramolacular CT. Although these endeavors are well in-line with our recent efforts devoted to local and Rydberg transitions 
of organic compounds \cite{Loo18a,Loo19c,Loo20a,Loo20d} that led to the QUEST database encompassing approximately 500 reference vertical transition energies, \cite{Loo20c,Ver21} it should be noted that the very nature of CT transitions makes the determination 
of reference values more challenging. Indeed, large density shifts ubiquitous to CT phenomena typically take place in larger compounds than those previously treated. Consequently, (EOM-)CCSDTQ calculations are clearly  beyond reach, whereas (EOM-)CCSDT 
calculations lie at the frontier of today's possibilities.  Beyond completing the QUEST database, \cite{Loo20c,Ver21} we also believe that the present reference values nicely complement the ones recently proposed by the Szalay group for intermolecular CT excitations. \cite{Koz20}
Following the philosophy of the QUEST database, we also wish to avoid any experimental input in order to avoid potential biases and ease theoretical cross comparisons.

%%%%%%%%%%%%%%%%%%%%
%%% METHODS %%%
%%%%%%%%%%%%%%%%%%%%
\section{Computational methods}

The investigated systems are displayed in Fig. \ref{Fig-1}. They include some of the previously described compounds (see Sec.~\ref{sec:CTlit}), as well as a significant series of new derivatives.

%--------------------------------------
\subsection{Geometries}
%--------------------------------------
Unless otherwise stated, we use CCSD(T) \cite{Pur82} or CC3  \cite{Chr95b,Koc95} to optimize the ground-state geometry of each compound. These optimizations are carried out with Dunning's {\TZ} basis set using 
CFOUR2.1 \cite{Mat20} that offers analytical GS nuclear gradients for both methods. As expected, we applied the most advanced approach, CC3, when possible, i.e., for the ``smallest'' systems investigated. The frozen-core 
(FC) approximation is enforced during these geometry optimizations. Cartesian coordinates and the corresponding optimization method are provided for each compound in the Supporting Information (SI). 

Spatial symmetry is enforced during the geometry optimization process which 
induces some constraints for specific molecules. For example, the $C_{2v}$ point group is enforced for aniline meaning that the amino group is planar. Experimentally, the \ce{NH2} group is puckered, but the symmetry constraint
allows for faster calculations as well as an easier ES tracking from one method to another.  Of course, such constraints might prevent direct comparisons with experiments, but it is 
well-known that vertical transition energies have no clear experimental equivalent anyway. \cite{Loo19b}

%--------------------------------------
\subsection{Basis sets}
%--------------------------------------
As further explained below, in a first stage, we perform ES calculations with Dunning's {\DZ} and {\TZ} basis sets applying systematically the FC approximation. This allows us to provide TBEs/{\TZ} reference values which are subsequently 
employed to benchmark wavefunction methods. There are several reasons for the choice of {\TZ}: i) the same basis set  was used in previous benchmark studies devoted to intramolecular CT, \cite{Pea08,Pea11,Cas19}  and ii) the addition 
of diffuse basis functions would yield lower Rydberg transitions and increase state mixing. This would be detrimental for identifying CT states in some derivatives (e.g., the peptides). We note that Kozma and coworkers went for an even 
more radical choice ({\DZ}) in their recent work, \cite{Koz20} but we acknowledge that the basis set effects are likely larger for the intramolecular cases treated here. Of course, the absence of diffuse functions in ES calculations is likely to 
result in (slightly) overestimated transition energies, \hl{which is why we also provide estimates with diffuse containing basis sets. This is also justified by the  different basis set dependencies of 
 wavefunction- and density-based methods.} \cite{Gin19,Loo19,Loo20e} \hl{Therefore,} in a second stage, we also \hl{perform CCSD calculations} \cite{Pur82,Scu87,Koc90b,Sta93,Sta93b} \hl{with
\emph{aug}-cc-pVTZ, as well as} CC2 \cite{Chr95} calculations with {\AVTZ} and {\AVQZ} so as to get \hl{estimates with larger basis sets. See below for further details.}

%--------------------------------------
\subsection{Reference calculations}
\label{sec:TBE}
%--------------------------------------
The first stage of the present study deals with the obtention of reference excitation energies for CT excited states. To identify CT transitions in the investigated derivatives, we first determine the lowest 8--20 transitions at the LR-CCSD/{\TZ} level with GAUSSIAN 16. 
\cite{Gaussian16}  We then analyze the nature of the  underlying orbitals, and, when possible, compare with literature results. Next, we compute the same ESs at both the ADC(2)/{\TZ} \cite{Tro97,Dre15} and CAM-B3LYP/{\TZ} \cite{Yan04} levels of theory using Q-CHEM 5.3 \cite{Kry13} 
and GAUSSIAN 16,  \cite{Gaussian16}  respectively. Establishing the correspondence between ESs at different levels of theory is straightforward for the vast majority of the cases. From the CAM-B3LYP calculations, we 
compute the CT distance, as given by Le Bahers' model \cite{Leb11c,Ada15} on the basis of the difference between the relaxed TD-DFT density and its GS KS-DFT counterpart. This value is simply labelled $d^{\text{CT}}_{\text{CAM}}$ below. 
Likewise, from the ADC(2) data, we compute the electron-hole distance from an analysis of the transition density matrix, \cite{Pla14,Pla14b} labelled $r^{\text{eh}}_{\text{ADC}}$ in the following.  Finally, we also compute, as an estimate of the CT strength, the
electron-hole distance determined from the inverse of the expectation value of the direct Coulomb operator over BSE electron-hole eigenstates stemming from the BSE/ev$GW$@HF/{\TZ} calculations (see also the SI). These are performed with the {\sc{Fiesta}} package. \cite{Jac15a}
These latter values are denoted $r^{\text{eh}}_{\text{BSE}}$ in the following. Whilst it would certainly be possible to rely on alternative metrics (see Sec.~\ref{sec:CTmet}), we have selected these three models to have complementary 
views on the nature of the CT states (DFT \emph{vs} wavefunction \emph{vs} Green's function, ES density \emph{vs} transition density \emph{vs} Coulomb matrix). As mentioned below, the CT transitions found following such a protocol are usually in agreement with the known literature.

Next, we use CFOUR, \cite{Mat20} to determine (EOM) CCSDT-3,  \cite{Wat96,Pro10} CC3, \cite{Chr95b,Koc95} and CCSDT \cite{Nog87,Scu88,Kuc01,Kow01,Kow01b} transition energies for the states previously identified. For the rather small number of pathological cases, having a LR-CCSD
guess is a valuable asset to ease the convergence towards the target ESs.  To define our TBEs/{\TZ} values, we rely on the following incremental approach
\begin{equation}\label{eq1}
\begin{split}
	\Delta E^{\text{TBE}}_{\text{TZ}} 
	& = \Delta E^{\text{CCSDT}}_{\text{cc-pVDZ}}
	+ \left[  \Delta E^{\text{CC3}}_{\text{cc-pVTZ}}   - \Delta E^{\text{CC3}}_{\text{cc-pVDZ}} \right]
	\\
	& = \Delta E^{\text{CCSDT}}_{\text{cc-pVDZ}} + \Delta\Delta E^{\text{CC3}}_{\text{TZ}},
\end{split}
\end{equation}
Such additive scheme is popular in the CC community \cite{Kal04,Bal06,Kam06b,Wat12,Fel14,Fra19,Chr21} and similar approaches have been employed in studies involving CT states. \cite{Pea12,Gui18b,Cas19} 
\hl{In Table S5 of the SI, we list the} $\Delta\Delta E_{\text{TZ}}$ \hl{values obtained with CCSDT-3 and CC3, and their very high degree of similarity is obvious, with a} $R^2$ \hl{of 0.99 and a mean absolute deviation between the two sets
of data as small as 0.01 eV.  In a second step, we have obtained TBEs accounting for diffuse orbitals by applying a similar scheme, that is,
}
\begin{equation}\label{eq2}
\begin{split}
	\Delta E^{\text{TBE}}_{\text{ATZ}} 
	& =\Delta E^{\text{TBE}}_{\text{TZ}}
	+ \left[  \Delta E^{\text{CCSD}}_{\text{aug-cc-pVTZ}}   - \Delta E^{\text{CCSD(T-3)}}_{\text{cc-pVTZ}} \right]
	\\
	& =\Delta E^{\text{TBE}}_{\text{TZ}} +  \Delta\Delta E^{\text{CCSD(T-3)}}_{\text{ATZ}}
	\\
	& = \Delta E^{\text{CCSDT}}_{\text{cc-pVDZ}} + \Delta\Delta E^{\text{CC3}}_{\text{TZ}} +  \Delta\Delta E^{\text{CCSD(T-3)}}_{\text{ATZ}}
\end{split}
\end{equation}
\hl{in which the term} $\Delta\Delta E_{\text{ATZ}}$ \hl{was typically determined with CCSD, unless CCSDT-3/\emph{aug}-cc-pVTZ calculations were technically feasible.  In Table S6 in the SI, we compare
the} $\Delta\Delta E_{\text{ATZ}}$ \hl{values obtained with CC2, CCSD, and CCSDT-3. Whilst the basis set corrections are highly dependent on the considered state and molecule, they are almost unaffected by the
level of theory selected, e.g., the absolute difference between the CCSD and CCSDT-3 basis set correction is at most 0.02 eV and 0.01 eV on average. This clearly highlights the transferability of basis set effects 
between these two wavefunction methods.} Eventually, to get \hl{even} closer to the CBS limit and ease the comparison between wavefunction- and density-based methods, we \hl{added additional corrections at the CC2 level, i.e.}
\begin{equation}\label{eq3}
\begin{split}
	\Delta E^{\text{TBE}}_{\text{AQZ}} 
	& = \Delta E^{\text{TBE}}_{\text{ATZ}} + \left[  \Delta E^{\text{CC2}}_{\text{aug-cc-pVQZ}} -  \Delta E^{\text{CC2}}_{\text{aug-cc-pVTZ}} \right] 
	\\
	& = \Delta E^{\text{TBE}}_{\text{ATZ}} + \Delta\Delta E^{\text{CC2}}_{\text{AQZ}} 
	\\
	& =  \Delta E^{\text{CCSDT}}_{\text{cc-pVDZ}} + \Delta\Delta E^{\text{CC3}}_{\text{TZ}} +  \Delta\Delta E^{\text{CCSD(T-3)}}_{\text{ATZ}} + \Delta\Delta E^{\text{CC2}}_{\text{AQZ}} 
\end{split}
\end{equation}
The CC2 calculations with both {\AVTZ} and {\AVQZ} are performed with TURBOMOLE, \cite{Turbomole} applying the resolution-of-identity (RI) approximation with the corresponding basis sets. \cite{Wei02} and we have confirmed that the
RI approximation has a negligible effect on the present results. \hl{As can be seen below, this last correction is marginal for the vast majority of states considered here, so that we do expect that the}
{\AVQZ} basis set provides very accurate estimates for low-lying ESs in organic compounds, and we do not foresee further basis set extension to play a significant role.

%--------------------------------------
\subsection{Wavefunction and BSE benchmarks}
%--------------------------------------

In the second phase of the present study, we evaluate the performances of several wavefunction- and Green's function-based approaches using the $\Delta E^{\text{TBE}}_{\text{TZ}}$ values defined in Sec.~\ref{sec:TBE} as references. 
We systematically apply the FC approximation in all these calculations.
The following approaches were tested: CIS(D),  \cite{Hea94,Hea95} EOM-MP2,  \cite{Sta95c}   \hl{SOPPA}, \cite{Nie80b,Bak00} \hl{RPA(D)}, \cite{Chr98e}
CC2,  \cite{Chr95,Hat00} CCSD, \cite{Pur82} CCSD(T)(a)*, \cite{Mat16} CCSDR(3),  \cite{Chr96b}  CCSDT-3, \cite{Wat96,Pro10} CC3,  \cite{Chr95b,Koc95}  ADC(2), \cite{Tro97,Dre15} ADC(3), \cite{Tro02,Har14,Dre15} 
ADC(2.5), \cite{Loo20b}  and BSE/$GW$. \cite{Sal51,Han79,Bla18} The EOM-MP2 and ADC calculations are performed with Q-CHEM 5.2, \cite{Kry13} applying the RI approximation with the cc-pVTZ-RI auxiliary basis set, \cite{Wei02}
and tightening the convergence and integral thresholds. The CIS(D) and CCSD calculations are achieved with GAUSSIAN 16, \cite{Gaussian16} using default parameters. The \hl{SOPPA, RPA(D),} CC2 and CCSDR(3) results are obtained with 
DALTON 2017 \cite{dalton}, also using default parameters. In the following, we omit the prefixes LR and EOM as both formalisms are known to yield identical excitation energies. \cite{Row68,Sta93} 

The BSE calculations  are performed with the {\sc{Fiesta}} package, \cite{Jac15a} using Coulomb-fitting RI with the cc-pVTZ-RI auxiliary basis set. \cite{Wei02} The intermediate $GW$ quasiparticle energies
and screened Coulomb potential $W$ are calculated using a partially self-consistent scheme on the eigenvalues (ev$GW$) shown in several studies to provide accurate data \cite{Kap16,Ran16c} while significantly removing the dependency on the input KS or HF
eigenstates in the final BSE excitation energies. \cite{Jac15a,Jac16c,Gui18b}  Dynamical effects in the $GW$ self-energy are treated within an exact contour-deformation approach. For good convergence, all MO energy levels within 10 eV of the HOMO-LUMO gap are explicitly 
corrected at the $GW$ level, lower (higher) states being shifted using the quasiparticle correction obtained for the lowest (highest) explicitly corrected level. \cite{Jac16c}  To facilitate the identification of transitions, we first focus on BSE/ev$GW$ calculations starting from HF 
eigenstates (BSE/ev$GW$@HF), but we next determine the BSE/ev$GW$ excitation energies obtained starting from PBE0 \cite{Ada99,Erz99} eigenstates (BSE/ev$GW$@PBE0), which is a more usual choice in BSE calculations.  

\subsection{TD-DFT benchmarks}

All our TD-DFT calculations have been performed with GAUSSIAN 16, \cite{Gaussian16} using the \emph{ultrafine} quadrature grid. As the convergence with respect to the basis set size of vertical excitation energies stemming from density-based methods (such as TD-DFT) 
and wavefunction-based methods tend to significantly differ, \cite{Loo19,Gin19,Loo20e} we have decided to perform the TD-DFT benchmarks with the {\AVQZ} basis set (i.e., using the TBE/{\AVQZ} values as references), which is likely large enough to be close to the CBS limit for both families of methods. 
We have selected the following XCFs to perform our calculations: two global hybrids with rather low exact exchange percentage, B3LYP (20\%) \cite{Bec93,Fri94,Bar94,Ste94} and PBE0 (25\%), \cite{Ada99,Erz99} one global hybrid with a much larger share of exact 
exchange, M06-2X (54\%), \cite{Zha08b} and five RSHs (CAM-B3LYP, \cite{Yan04} LC-$\omega$HPBE, \cite{Hen09c} $\omega$B97X, \cite{Cha08} $\omega$B97X-D, \cite{Cha08b} and M11 \cite{Pev11c}). As mentioned in Sec.~\ref{sec:CTex}, it is
well recognized that the latter XCFs are better suited for modeling CT transitions. We wish nevertheless to explore the performances of the global hybrids for ``mild'' CT as well as the relative performances of the five RSH functionals for various CT strengths.

%%%%%%%%%%%%%%%%%%%%
%%% RESULTS %%%
%%%%%%%%%%%%%%%%%%%%

%%% TABLE I %%%
\begin{sidewaystable*}
\caption{Reference data for CT ESs. For each state, we provide its symmetry and three CT parameters (in \AA, see Sec.~\ref{sec:CTmet}), as well as the transition energies (in eV) obtained with various wave function methods,
the TBE/{\TZ} and TBE/{\AVTZ}  reference excitation energies, the CC2 correction $\Delta\Delta E^{\text{CC2}}$, and the corresponding basis-set corrected TBE/{\AVQZ} values. These values are obtained with Eqs.~\ref{eq1} to \ref{eq3}, 
except otherwise stated. Comparisons with literature are given in the rightmost columns.} 
\label{Table-1}
\footnotesize
\vspace{-0.3 cm}
\begin{tabular}{p{3.5cm}l|ccc|cccc|cccc|ccc|cc|cccc}
\hline 
		&								& \multicolumn{3}{c}{{\TZ}}	&\multicolumn{4}{c}{{\DZ}} & \multicolumn{4}{c}{{\TZ}}		& \multicolumn{3}{c}{{\AVTZ}}							&\multicolumn{2}{c}{{\AVQZ}}		& \multicolumn{4}{c}{Litt.}   \\
Molecule 	& State					&$d^{\text{CT}}_{\text{CAM}}$&$r^{\text{eh}}_{\text{ADC}}$ &$r^{\text{eh}}_{\text{BSE}}$
														&CCSD	&CCSDT-3 	& CC3	& CCSDT		&CCSD	&CCSDT-3 	& CC3	& TBE		&\hl{CCSD}&\hl{CCSDT-3}&\hl{TBE}	&$\Delta\Delta E^{\text{CC2}}_{\text{AQZ}} $	&TBE	& Th.	& Th.	&Exp.	&Exp.\\
\hline
Aminobenzonitrile&2 $A_1$ ($\pi \rightarrow \pi^\star$)&1.15&1.01&2.05	&5.53	&5.43		&5.39	&5.39		&5.41	&5.30		&5.25	&\textbf{5.26}	&\hl{5.23}	&\hl{5.13}		&\textbf{5.09}	&\hl{0.00}	&\textbf{5.09}	&4.98$^a$&5.13$^b$&4.76$^c$\\
Aniline	&2 $A_1$ ($\pi \rightarrow \pi^\star$)		&1.02&0.83&1.91	&6.16	&6.08		&6.04	&6.05		&5.99	&5.90		&5.86	&\textbf{5.87}	&\hl{5.60}	&\hl{5.53}		&\textbf{5.50}	&\hl{-0.02}	&\textbf{5.48}	&5.42$^d$&5.34$^e$&5.39$^f$&\\
Azulene	&2 $A_1$ ($\pi \rightarrow \pi^\star$)		&1.16&1.06&2.36	&4.12	&4.01		&3.98	&3.99		&4.02	&3.92		&3.88	&\textbf{3.89}	&\hl{3.97}	&			&\textbf{3.85}	&\hl{0.00}	&\textbf{3.84}	&3.81$^g$&3.46$^h$&3.56$^i$&3.57$^j$\\
		&2 $B_2$ ($\pi \rightarrow \pi^\star$)		&1.02&0.95&2.43	&4.89	&4.68		&4.60	&4.62		&4.82	&4.61		&4.52	&\textbf{4.55}	&\hl{4.78}	&			&\textbf{4.50}	&\hl{-0.01}	&\textbf{4.49}	&4.15$^g$&4.13$^h$&4.23$^k$\\
Benzonitrile&1 $A_2$ ($\pi_{\text{CN}} \rightarrow \pi^\star$)
										&1.17&1.18&1.73	&7.48	&7.31		&7.25	&7.27		&7.33	&7.15		&7.08	&\textbf{7.10}	&\hl{7.28}	&\hl{7.10}		&\textbf{7.05}	&\hl{0.00}	&\textbf{7.05}	&7.37$^a$\\
Benzothiadiazole&1 $B_2$ ($\pi \rightarrow \pi^\star$)&1.41&1.24&2.30	&4.82	&4.59		&4.50	&4.56		&4.63	&4.40		&4.30	&\textbf{4.37}	&\hl{4.56}	&\hl{4.32}		&\textbf{4.29}	&\hl{-0.01}	&\textbf{4.28}	&		&	&3.52$^l$	\\
Dimethylaminobenzonitrile	
		&2 $A_1$ ($\pi \rightarrow \pi^\star$)		&1.48&1.44&2.15	&5.20	&5.10		&5.05	&5.06		&5.10	&4.99		&4.93	&\textbf{4.94}	&\hl{5.02}	&			&\textbf{4.86}	&\hl{0.00}	&\textbf{4.86}	&4.90$^m$&4.94$^n$&4.57$^o$&\\
Dimethylaniline&1 $B_2$ ($\pi \rightarrow \pi^\star$)	&1.13&0.98&2.20	&4.74	&4.63		&4.59	&4.58		&4.66	&4.53		&4.48	&\textbf{4.47}	&\hl{4.58}	&			&\textbf{4.39}	&\hl{0.00}	&\textbf{4.40}	&4.30$^p$&4.48$^q$&4.30$^r$\\
			&2 $A_1$ ($\pi \rightarrow \pi^\star$)	&1.25&1.22&2.02	&5.81	&5.73		&5.68	&5.69		&5.68	&5.58		&5.53	&\textbf{5.54}	&\hl{5.54}	&			&\textbf{5.40}	&\hl{0.00}	&\textbf{5.40}	&5.06$^p$&		&5.16$^r$\\
Dipeptide	%&$A'$	($\pi_1 \rightarrow \pi_2^\star$)	&&				&7.50	&7.36		&7.29	&7.37		&7.35	&7.19		&7.11	&\textbf{7.19}	&		&			&			&		&			&7.18$^a$	&7.17$^b$\\%UNCLEAR STATE-REMOVE
		& 7 $A''$	($n_1 \rightarrow \pi_2^\star$)	&2.17&3.62&3.35	&9.07	&8.53		&8.28	&8.39		&8.92	&8.31		&8.04	&\textbf{8.15}	&		&			&			&		&			&8.07$^s$	&8.33$^t$\\
$\beta$-Dipetide&7 $A'$ ($\pi_1 \rightarrow \pi_2^\star$)&2.36&3.16	&3.11&9.13	&8.85		&8.72	&8.77		&8.90	&8.59		&		&\textbf{8.51}$^u$&		&			&			&		&			&7.99$^s$	&8.59$^t$\\
		&10 $A''$($n_1 \rightarrow \pi_2^\star$)	&2.29&4.35&3.22	&9.83	&9.32		&9.08	&9.20		&9.58	&9.02		&		&\textbf{8.90}$^u$&		&			&			&		&			&9.13$^s$	&9.08$^t$\\
Hydrogen Chloride&1 $\Pi$ ($n  \rightarrow \sigma^\star$)	&1.05&0.95&1.66&8.29	&8.24		&8.23	&8.23		&8.18	&8.12		&8.11	&\textbf{8.10}$^v$&\hl{7.91}&\hl{7.85}	&\textbf{7.84}$^v$&	&\textbf{7.88}$^w$&7.86$^x$&8.23$^y$\\
Nitroaniline	& 2 $A_1$ ($\pi \rightarrow \pi^\star$)&2.02&2.08&2.27	&4.96	&4.79		&4.70	&4.76		&4.80	&4.61		&4.51	&\textbf{4.57}	&\hl{4.63}	&			&\textbf{4.40}	&\hl{-0.01}	&\textbf{4.39}	&4.54$^t$&4.30$^{z}$\\
Nitrobenzene& 2 $A_1$ ($\pi \rightarrow \pi^\star$)	&1.66&1.51&2.07	&6.00	&5.84		&5.78	&5.83		&5.77	&5.59		&5.52	&\textbf{5.57}	&\hl{5.62}	&\hl{5.43}		&\textbf{5.41}	&\hl{-0.01}&\textbf{5.39}&4.99$^{aa}$&5.27$^{ab}$&4.62$^{ac}$&5.11$^{ad}$\\
Nitrodimethylaniline&2 $A_1$ ($\pi \rightarrow \pi^\star$)&2.18&2.41	&2.37&4.68	&4.51		&4.42	&4.48		&4.53	&4.33		&4.22	&\textbf{4.28}	&\hl{4.39}	&			&\textbf{4.14}	&\hl{-0.01}	&\textbf{4.13}	&		&		&3.89$^{ae}$\\
Nitropyridine N-Oxide&2 $A_1$ ($\pi \rightarrow \pi^\star$)&1.70&1.97&2.07&4.60	&4.39		&4.28	&			&4.46	&4.24		&4.13	&\textbf{4.24}$^{af}$&\hl{4.32}&\hl{4.10}	&\textbf{4.10}	&\hl{0.00}	&\textbf{4.10}	&4.32$^{ag}$	&		&3.80$^{ah}$\\		
N-Phenylpyrrole&2 $B_2$  ($\pi \rightarrow \pi^\star$)&2.11&2.13&2.07	&6.02	&5.77		&5.67	&5.70		&5.84	&5.60		&5.50	&\textbf{5.53}	&\hl{5.63}	&			&\textbf{5.32}	&\hl{0.00}	&\textbf{5.32}	&5.52$^t$&5.21$^{ai}$& \\
			&3 $A_1$ ($\pi \rightarrow \pi^\star$)	&2.28&3.54&3.54	&6.72	&6.33		&6.17	&6.24		&6.52	&6.14		&5.97	&\textbf{6.04}	&\hl{6.34}	&			&\textbf{5.85}	&\hl{0.00}	&\textbf{5.86}	&6.07$^t$&5.69$^{ai}$& \\
Phthalazine	&1 $A_2$ ($n \rightarrow \pi^\star$)	&1.11&1.87&2.00	&4.24	&4.05		&3.92	&3.95		&4.26	&4.03		&3.89	&\textbf{3.93}	&\hl{4.25}	&\hl{4.01}		&\textbf{3.91}	&\hl{0.01}	&\textbf{3.91}	&3.74$^{aj}$&3.68$^{ak}$ &3.61$^{al}$&3.01$^{am}$	\\
			&1 $B_1$ ($n \rightarrow \pi^\star$)	&1.13&1.87&1.86	&4.67	&4.49		&4.38	&4.40		&4.64	&4.43		&4.32	&\textbf{4.34}	&\hl{4.61}	&\hl{4.40}		&\textbf{4.31}	&\hl{0.00}	&\textbf{4.31}	&4.20$^{aj}$&4.12$^{ak}$ &3.91$^{al}$&3.72$^{am}$	\\
Quinoxaline	&1 $B_2$ ($\pi \rightarrow \pi^\star$)	&1.51&1.76&2.42	&5.20	&4.99		&4.90	&4.95		&5.00	&4.79		&4.69	&\textbf{4.74}	&\hl{4.91}	&\hl{4.69}		&\textbf{4.64}	&\hl{-0.01}&\textbf{4.63}	&4.45$^{aj}$&4.20$^{ak}$&4.34$^{al}$&3.96$^{an}$\\
			&3  $A_1$ ($\pi \rightarrow \pi^\star$)&1.12&0.97&2.18	&6.13	&5.97		&5.90	&5.89		&6.01	&5.84		&5.76	&\textbf{5.75}	&\hl{5.91}	&\hl{5.75}		&\textbf{5.66}	&\hl{-0.01}	&\textbf{5.65}	&		&		&5.36$^{al}$&5.36$^{an}$\\
			&2 $B_1$ ($n \rightarrow \pi^\star$)	&1.21&1.85&2.30	&6.94	&6.59		&6.39	&6.46		&6.87	&6.48		&6.26	&\textbf{6.33}	&\hl{6.73}	&\hl{6.36}		&\textbf{6.21}	&\hl{0.01}	&\textbf{6.22}	&		&			&		&		\\
Twisted DMABN&1 $A_2$($n \rightarrow \pi^\star$)	&1.99&2.69&2.64	&4.43	&4.31		&4.23	&4.24		&4.41	&4.25		&4.15	&\textbf{4.17}	&\hl{4.35}	&			&\textbf{4.11}	&\hl{0.01}	&\textbf{4.12}	&4.25$^{ao}$&\\%-0.055
			&1 $B_1$($n \rightarrow \pi^\star$)	&1.74&2.60&2.17	&5.27	&5.07		&4.95	&4.98		&5.19	&4.95		&4.81	&\textbf{4.84}	&\hl{5.09}	&			&\textbf{4.74}	&\hl{0.01}	&\textbf{4.75}	&5.09$^{ao}$&\\%-0.085
Twisted PP	&2 $B_2$  ($\pi \rightarrow \pi^\star$)&2.33&3.32&3.51	&6.19	&5.91		&5.79	&5.85		&6.10	&5.80		&5.67	&\textbf{5.73}	&\hl{5.95}	&			&\textbf{5.58}	&\hl{0.00}	&\textbf{5.58}	&5.79$^{ao}$&5.35$^{ap}$\\
			&2 $A_1$  ($\pi \rightarrow \pi^\star$)&2.38&3.32&2.96	&6.37	&6.09 		&5.97	&6.03		&6.18	&5.89		&5.76	&\textbf{5.82}	&\hl{6.00}	&			&\textbf{5.64}	&\hl{0.00}	&\textbf{5.65}	&		     &5.45$^{ap}$\\
			&1 $A_2$  ($\pi \rightarrow \pi^\star$)&2.32&3.27&3.07	&6.41	&6.21		&6.11	&6.14		&6.35	&6.12		&6.01	&\textbf{6.04}	&\hl{6.26}	&			&\textbf{5.95}	&\hl{0.01}	&\textbf{5.95}	&		     &5.89$^{ap}$\\
			&1 $B_1$  ($\pi \rightarrow \pi^\star$)&2.37&3.37&3.15	&6.78	&6.54		&6.42	&6.46		&6.61	&6.36		&6.24	&\textbf{6.28}	&\hl{6.50}	&			&\textbf{6.16}	&\hl{0.01}	&\textbf{6.17}	&6.31$^{ao}$ &\\
\hline	
\end{tabular}
\vspace{-0.3 cm}
\begin{flushleft}
\begin{scriptsize}
$^a${CASPT2/DZP value from Ref.~\citenum{Sob96};}
$^b${STEOM-CC/{\DZ} value from Ref.~\citenum{Par99};}
$^c${Experimental maximum in $n$-heptane from Ref.~\citenum{Zac93};}
$^d${CR-EOM-CCSD(T)/{\AVDZ} value from Ref.~\citenum{Wan13e}};
$^e${SAC-CI value from Ref.~\citenum{Hon02b};}
$^f${$\lambda_{\text{max}}$ (vapor phase) from Ref.~\citenum{Kim64};}
$^g${Vertical CASPT2/6-31G(d) results from Ref.~\citenum{Mur04b};}
$^h${0-0 DFT(BHHLYP)/MRCI/TZVPP values from Ref.~\citenum{Vos15};}
$^i${0-0 energy from fluorescence study of Ref.~\citenum{Hir78};}
$^j${Photoelectron spectroscopy from Ref.~\citenum{Vos15};}
$^k${0-0 energy from the fluorescence spectrum of the jet-cooled derivative in Ref.~\citenum{Fuj83};}
$^l${0-0 energy measured in frozen dichlorobenzene matrix from Ref.~\citenum{Lin78};}
$^m${MRCIS(8,7)+P/ANO-DZ value from Ref.~\citenum{Geo15};}
$^n${ADC(3)/{\DZ} result from Ref.~\cite{Mew17};}
$^o${Experimental $\lambda_{\text{max}}$ (vapor phase) from Ref.~\citenum{Dru10};}
$^p${CASPT2/6-311G(d,p) values from Ref.~\citenum{Gal09};}
$^q${CCSDR(3)/{\AVDZ} result from Ref.~\citenum{Tho15};}
$^r${Experimental $\lambda_{\text{max}}$ (vapour phase) from Ref.~\citenum{Kim64};}
$^s${CASPT2/DZP value from Ref.~\citenum{Ser98b};}
$^t${CCSDR(3)/{\TZ} value (basis set extrapolated for the $\beta$-Dipeptide) from Ref.~\citenum{Cas19};}
$^u${CCSDT-3/{\TZ} value corrected by the difference between CCSDT/{\DZ} and CCSDT-3/{\DZ} energies;}
$^v${FCI/{\TZ} value (present paper) \hl{and FCI/}{\AVTZ} \hl{from Ref}.~\citenum{Loo18a}. For the former basis, the same result is obtained with CCSDTQ/{\TZ};}
$^w${Present CCSDTQ/{\AVQZ} value.}
$^x${FCI/CBS (no FC) estimate from Ref.~\citenum{Loo18a};}
$^y${CC2/{\TZ} value from Ref.~\citenum{Pea08};}
$^{z}${$\gamma$-CR-EOMCC(2,3),D/6-31+G(d,p) value from Ref.~\citenum{Hoy16};}
$^{aa}${CASPT2//B3LYP result from Ref.~\citenum{Kro00}, the most recent CASPT2 we are aware of reports a similar value of 5.01 eV \cite{Giu17};}
$^{ab}${ADC(3)/\emph{def2}-TZVP energy from Ref.~\citenum{Mew14};}
$^{ac}${EELS on monolayer coverage from Ref.~\citenum{Kro00};}
$^{ad}${Gas-phase optical absorption maximum from Ref.~\citenum{Nag64}, a similar value of 5.15 eV is reported in the more recent Ref.~\citenum{Kri16};}
$^{ae}${Experimental gas phase value from Ref.~\citenum{Lau94};}
$^{af}${CCSDT-3/{\TZ} value, see text;} 
$^{ag}${CASPT2/{\AVDZ} from Ref.~\citenum{Bud16b} (previously unpublished);}
$^{ah}${Extrapolated gas-phase maximum, see Ref.~\citenum{Bud16b};}
$^{ai}${Ext. CC3/{\AVTZ} value from Ref.~\citenum{Gui18b};}
$^{aj}${CC2/{\AVDZ} results from Ref.~\citenum{Eti17}:}
$^{ak}${CASPT2/{\QZ} results from Ref.~\citenum{Mor09};}
$^{al}${From MCD spectra in $n$-heptane from Ref.~\citenum{Kai78};}
$^{am}${0-0 energy from Ref.~\citenum{Inn88};}
$^{an}${0-0 energy in vapor from Ref.~\citenum{Gla70};}
$^{ao}${Ext. CCSD/{\AVTZ} values from Ref.~\citenum{Pea12};}
$^{ap}${CASPT2/ANO-DZP values from Ref.~\citenum{Pro00}.}
\end{scriptsize}
\end{flushleft}
\end{sidewaystable*}
%%% %%% %%% %%%

\section{Results and Discussion}

\subsection{Reference values}

Our reference vertical excitation energies are listed in Table \ref{Table-1}, in which we report CCSD, CCSDT-3, CC3, and CCSDT values with two basis sets ({\DZ} and {\TZ}), as well as literature values and CT strengths evaluated thanks to the three models described in Sec.~\ref{sec:TBE}.  
Additional details (oscillator strengths,  MO combinations at CCSD level, etc) can be found in the SI. Taking the $r^{\text{eh}}_{\text{ADC}}$  values as reference, one notes a satisfactory agreement with $d^{\text{CT}}_{\text{CAM}}$ for rather small CT (for strong CT, the CAM-B3LYP charge 
separations appear too small), and a decent match with $r^{\text{eh}}_{\text{BSE}}$ for the cases of large electron-hole separation (for weak CT, the BSE charge separations appear too large). This can be clearly seen in Fig.~S1 in the SI.  At this stage, we of course highlight that in
the case of mild CT character, a change of basis set could induce non-trifling variations of the values given by these CT metrics.

\subsubsection{Aminobenzonitrile}

Aminobenzonitrile (ABN) is a well-known push-pull molecule, that has been the subject to several previous theoretical studies with wavefunction approaches, \cite{Ser95,Sob96,Ser97b,Par99,Gom15,Seg16,Cas18} these works typically 
focussing on the two lowest ESs of local  ($B_2$) and CT ($A_1$) character. According to the ADC(2) metrics, excitation to this $A_1$ state induces a charge separation of ca.~$1$ \AA\ (see Table \ref{Table-1}), and there is a perfect match between the 
CC3 and CCSDT values, whereas CCSDT-3 (CCSD) seem to deliver slightly (significantly) overestimated values. Our TBE/{\TZ}, $5.26$ eV,  perfectly matches the value obtained in the most recent CASPT2 study we are aware of, \cite{Seg16} 
and is also in quite good agreement with a twenty-year-old STEOM-CCSD estimate ($5.13$ eV). \cite{Par99} Those two works used the {\DZ} basis set however. In contrast, all previous CASPT2 estimates seem to provide smaller values in the 
$4.44$--$5.01$ eV range. \cite{Ser95,Sob96,Ser97b,Gom15}  The experimental $\lambda_{\text{max}}$ is located at $4.76$ eV in an apolar solvent, \cite{Zac93} a value significantly below our basis set corrected vertical  transition energy ($5.09$ eV),
as expected in such comparison.

\subsubsection{Aniline}

For aniline, the lowest $A_1$ excited state involves more than one MO pair (see the SI) and has a weak CT character ($d^{\text{CT}}_{\text{CAM}} = 1.02$ \AA, $r^{\text{eh}}_{\text{ADC}} = 0.83$ \AA, and $r^{\text{eh}}_{\text{BSE}} = 1.91$ \AA). 
Indeed, this state was previously characterized as local in a calculation involving the puckered amine, \cite{Hon02b} whereas we enforced the $C_{2v}$ point group in the present calculations. The results listed in Table \ref{Table-1} show a remarkable methodological 
stability, the CC3, CCSDT-3, and CCSDT values all falling inside a tight $0.04$ eV window with the {\DZ} basis set, whereas the differences obtained with the triple-$\zeta$ basis set are relatively small. Our TBE/{\TZ} value, $5.87$ eV, is
strongly lowered when further basis set corrections are accounted for ($5.48$ eV). The latter value is in good agreement with the investigation of Worth's group \cite{Wan13e} although we recall that we have enforced $C_{2v}$ symmetry here (which also makes comparisons with experiment difficult).

\subsubsection{Azulene}

Azulene is a very well-known asymmetric isomer of naphthalene. Its electronic transitions have been investigated at various levels of theory, \cite{Die04,Mur04b,Vos15,Vey20} likely due to its unusual non-Kasha fluorescence.
According to the considered metrics, the second ($2 A_1$) and third ($2 B_2$) singlet ESs exhibit small CT characters. As can be seen in Table \ref{Table-1}, CC3 transition energies are again very close from the CCSDT ones,
whereas going from the double- to the triple-$\zeta$ basis set decreases the predicted transition energies by roughly $-0.10$ eV, further basis set extensions yielding even smaller changes.  Our TBE is very close to a previous 
CASPT2/6-31G(d) estimate \cite{Mur04b} for the $A_1$ ES, but is significantly higher than the multi-reference result for the $B_2$ ES.  Both our TBEs exceed the experimental 0-0 energies by approximately $0.3$--$0.4$ eV, \cite{Hir78,Fuj83,Vos15} 
which is the expected trend.

\subsubsection{Benzonitrile}

In benzonitrile the two lowest transitions of $A_1$ and $B_2$ symmetries do not present any significant CT character (not shown). There is however a higher-lying dark  $A_2$ state corresponding to a CT from the 
$\pi$ orbital of the cyano moiety parallel to the main molecular plane towards the highly-delocalized LUMO (see the SI for representation of the MOs) that has a CT nature ($r^{\text{eh}}_{\text{ADC}} = 1.18$ \AA). 
Our TBE/{\TZ} for this transition is $7.10$ eV, which is likely trustworthy as the CC3 and CCSDT results are much alike with the {\DZ} basis set (see Table \ref{Table-1}). To  
the best of our knowledge, this specific transition was not investigated previously, but for a rather old CASPT2 analysis that reports a $7.33$ eV value for the lowest $A_2$ ES. \cite{Sob96}

\subsubsection{Benzothiadiazole}

This bicyclic system, BTD, is an extremely popular acceptor unit in solar cell applications. \cite{Par09b,Wu13,Mat14} Surprisingly, while one can find many TD-DFT investigations of large dyes encompassing a BTD moiety, there seems to be no previous
wavefunction investigation of this (isolated) building block. Contrasting with the previous molecules, the CCSDT-3 transition energy is closer from the CCSDT value than  its CC3 counterpart, though all three methods provide very similar excitation energies.
Our TBE/{\AVQZ} of $4.28$ eV, is $0.76$ eV above the experimental 0-0 energy, \cite{Lin78} but such large value is not inconsistent with the large experimental Stokes shift, \cite{Net12} and the strong theoretical elongation of the \ce{N-S} 
bonds in going from the GS to the ES. \cite{Lau14b}

\subsubsection{Dimethylaminobenzonitrile}

Dimethylaminobenzonitrile (DMABN) is the prototypical system undergoing twisted intramolecular CT (TICT) and it consequently displays a dual-fluorescence signature strongly dependent on the medium. This process has 
been the subject of countless investigations, \cite{Zac93,Ser95,Sob96,Roo96,Roo97,Ser97b,Par99,Hon02b,Koh04,Gri04b,Gri07,Rhe07,Hel08,Wig09,Rhe09,Ngu10,Ngu11,Mar11,Lun13,Hed13,Yan14b,Geo15,Gom15,Mew17,Mew18,Car19} 
and it is clearly not our intend to review in details all these works. Besides its TICT feature, DMABN is undoubtedly  one of the most popular dye  in CT benchmarks. \cite{Pea08,Pea12,Dev12,Hoy16,Jac17b,Gui18b,Cas19} 
However, the present work stands again  as the first to propose an estimate of CCSDT quality. Our TBE/{\AVQZ}, $4.86$ eV, should be rather solid given the agreement between CC3 and CCSDT, and the rather limited
basis set effects. This TBE is exactly the same as the extrapolated CC3/{\AVTZ} result of Ref.~\citenum{Gui18b} and is also close to the  ADC(3)/{\DZ} value given by Mewes and coworkers ($4.94$ eV). \cite{Mew17}
Of course, one can also found other estimates at lower levels of theory, e.g., $4.88$ eV with  CCSD/{\AVDZ}, \cite{Car19} $4.73$ eV with STEOM-CCSD, \cite{Par99} and 4.90 eV with
MRCIS(8,7)+P/ANO-DZ, \cite{Geo15} all three being reasonably close to the current TBE. In contrast, previous CASPT2 estimates of $4.41$ eV, \cite{Ser95} $4.51$ eV, \cite{Sob96} $4.47$ eV, \cite{Ser97b} $4.45$ eV \cite{Gom15} 
are all significantly too low.

\subsubsection{Dimethylaniline}

This derivative was much less investigated than the previous one and we could find only two studies of its excited states involving high-level \textit{ab initio} methods [CASPT2 \cite{Gal09} and CCSDR(3) \cite{Tho15}]. The metrics selected in this work describe the two lowest ESs 
of this compound as having a small CT character with a charge separation of ca.~1 \AA\ with ADC(2), the CT nature of the $A_1$ transition being only slightly larger than that of the ``local'' $B_2$ excitation. Here again, one notes the usual methodological trends as illustrated by the data gathered in 
Table \ref{Table-1}, with superb agreement between the CC3 and CCSDT estimates, and a limited drop of the transition energies when enlarging the basis set. The basis set corrected TBEs of $4.40$ and $5.40$ eV are reasonably in line with the literature. \cite{Gal09,Tho15}

\subsubsection{Dipeptide and $\beta$-dipeptide}

Both of these model compounds were originally characterized at the CASPT2 level by Serrano-Andr\'es and F\"ulscher. \cite{Ser98b} The smaller derivative is a popular test molecule for CT,  \cite{Roc10,Fab13} and is part of Tozer's \cite{Pea08,Pea12} 
and Goerigk's \cite{Cas19} sets. In both systems, these works identified two CT transitions, denoted as $\pi_1 \rightarrow \pi_2^\star$  and  $n_1 \rightarrow \pi_2^\star$,  the subscript referring to the amide number in the compound (see Fig.~\ref{Fig-1}).
For the smaller dipeptide,  Tozer relied on the CASPT2/ANO-"DZP" values of $7.18$ eV and $8.07$ eV as benchmarks in his original work, \cite{Pea08} whereas  Goerigk proposed reference values of $7.17$ and $8.33$ eV estimates obtained the CCSDR(3)/cc-pVTZ level. 

If these values seem numerically consistent, these two transitions are hardly well defined, the mixing of the MO character making unambiguous assignments impossible. As we detail in the SI, this is especially the case for the former 
$\pi_1 \rightarrow \pi_2^\star$ transition that mixes with local excitations. In fact at the same CCSD/{\TZ} level, Tozer selected the $8.09$ eV transition as a CT ES and the $7.35$ eV transition as a local ES, \cite{Pea12} whereas Goergik made the opposite assignments. Both choices
are in fact reasonable based on the selected criteria (see the SI). Due to this confusion, we did only consider the less problematic $n_1 \rightarrow \pi_2^\star$ excitation for the dipeptide. For this transition, the CCSDT value is interestingly in-between the
CC3 and CCSDT-3 estimates, rather than closer to the CC3 value. As such transition has, chemically speaking, an intermolecular nature, this outcome parallels the finding of Kozma and coworkers who found that CCSDT-3 performs better for intermolecular CTs. \cite{Koz20}
Our TBE/{\TZ} of $8.15$ eV is slightly smaller (larger) than previous CCSDR(3)/TZ (CASPT2/DZ) estimates. 

For the $\beta$-dipeptide, the identification of the two CT transitions is somehow easier than in the sister compound (see the SI for details).  With the {\DZ} basis, CCSDT values are roughly midway to CC3 and CCSDT-3. However, the larger size and limited symmetry 
of $\beta$-dipeptide make calculations extremely challenging, and CC3/{\TZ} calculations
were beyond our computational reach. For the  $\pi_1 \rightarrow \pi_2^\star$  excitation, the CCSDT value is bracketed by the CC3 and CCSDT-3 results, and our TBE of $8.51$ eV is slightly below the CCSDR(3) data of Ref.~\citenum{Cas19}, whereas the 
original CASPT2 transition energy is significantly too small.  \cite{Ser98b}  For the higher lying $n_1 \rightarrow \pi_2^\star$  excitation, our best estimate obtained with the same protocol is 8.90 eV, lies below previous estimates. \cite{Ser98b,Cas19} 

Given the very large MO mixing with both {\DZ} and {\TZ}, we did not attempt to obtain a TBE with diffuse-containing basis sets for these two derivatives.

\subsubsection{Hydrogen chloride}

HCl is small enough for allowing FCI and CCSDTQ calculations, and both yield a transition energy of 8.10 eV for the hallmark CT excitation with the {\TZ} basis set. This value is almost perfectly reproduced
by both CC3 and CCSDT-3.  We could also perform the CCSDTQ/{\AVQZ} calculation which returned an excitation energy of $7.88$ eV, within $0.02$ eV of our previous  FCI/CBS  value obtained on the
same geometry, but with a different computational strategy. \cite{Loo18a} Given these results, the original CC2/{\TZ} reference value considered in Tozer's set ($8.23$ eV) seems too large by $0.13$ eV, whereas the CC3/{\AVTZ} value of $7.81$ eV used in Ref.~\citenum{Gui18b} could be slightly too low.

\subsubsection{Nitroaniline}

pNA is a prototypical donor-acceptor system, the potent nitro group allowing an electron-hole separation of the order of $2$ \AA, about twice the distance determined in the related ABN compound.  As in the other nitro-bearing systems discussed below, 
CCSDT-3 seems to slightly outperform CC3, though the consistency of all CC approaches including triples remains excellent. Our TBEs are $4.57$ eV (with {\TZ}) and $4.39$ eV (with {\AVQZ}). This latter value is once more exactly equivalent to the 
one reported by the Klopper group with an extrapolated CC3/{\AVTZ} scheme. \cite{Gui18b} Other wavefunction estimates include a $3.80$ eV estimate with CASPT2, \cite{Ser97b} $4.30$ eV with  $\gamma$-CR-EOMCC(2,3)D/6-31+G(d,p), \cite{Hoy16} $4.72$ eV with 
EOM-CCSD/{\AVDZ}, \cite{Lu18}  and $4.54$ eV with an extrapolated CCSDR(3)/{\TZ} scheme. \cite{Cas19} 

\subsubsection{Nitrobenzene}

Similarly to the previous case,  in nitrobenzene, the pulling group is stronger than in benzonitrile, and the lowest $A_1$ state gains a significant CT character ($d^{\text{CT}}_{\text{CAM}} = 1.66$ \AA, $r^{\text{eh}}_{\text{ADC}} 
= 1.51$ \AA, and $r^{\text{eh}}_{\text{BSE}} = 2.07$ \AA). As for the other systems treated herein, the interested reader can find several previous calculations of the ES properties of this substituted system, \cite{Kro00,And08,Que11,Mew14,Kri16,Giu17,Sch18b} 
but to the best of our knowledge, none relied on a CC approach including contributions from the triples. Nevertheless, we wish to point out the joint exhaustive work by the Marian and Dreuw groups exploring the photophysics of 
nitrobenzene, \cite{Mew14} which includes CCSD, NEVPT2, and ADC(3) values for many ESs.  While the CCSD transition energy is, as expected too large, there is an excellent agreement between CCSDT and the 
other CC methods including iterative triples. The obtained TBEs are likely safe. These TBEs significantly exceed the experimental values, as well as the CASPT2  \cite{Kro00,Giu17} and ADC(3) estimates, \cite{Mew14} but are
 in good agreement with a recent CR-EOM-CCSD(T)/{\DZ} estimate of $5.44$ eV. \cite{Sch18b}

\subsubsection{Nitrodimethylaniline}

This chemical compound is likely one of the strongest donor-acceptor phenyl derivatives that one could envisage. The electron-hole separation in the lowest $A_1$ ES is enhanced by $0.1$--$0.3$ \AA\ and its energy is downshifted by roughly $-0.3$ eV 
 as compared to pNA. Otherwise, the methodological trends are exactly the same as in the parent compound, the CCSDT result being bracketed by the CCSDT-3 and CC3 values, and the basis set effects being within
 expectation for a low-lying ES.  Our TBE/{\AVQZ} is $0.24$ eV larger than the ``experimental'' $\lambda_{\text{max}}$ in gas phase, \cite{Lau94}  whereas we did not found previous CC estimates for this derivative.

\subsubsection{Nitropyridine N-Oxide}%DJ STOP (plus haut=relu et corrigŽ les petites variations)

This molecule is a solvatochromic probe, \cite{Lag96} and its interactions with various solvents were studied in details by various theoretical approaches. \cite{Bud16b} Besides it was not investigated theoretically as far as we know,
so that this is the first work reporting CC3 and CCSDT-3 transition energies. Unfortunately, the CCSDT/{\DZ} corrections failed to properly converge for that specific compound. As a consequence, given the results obtained for the
three previous nitro dyes, we went for the CCSDT-3/{\TZ} result as reference value. Our TBE/{\AVQZ} estimates are \hl{4.10} eV, \hl{0.30} eV above the gas-phase $\lambda_{\text{max}}$ value estimated in Ref.~\citenum{Bud16b} on the basis
of the experimental spectra of Ref.~\citenum{Lag96}. Again, such a difference between a vertical transition energy and an experimental absorption maximum is within expectations for a rigid dye.

\subsubsection{N-Phenylpyrrole}

PP is a well-known test molecules that is included in many CT sets. \cite{Pro00,Xu06,Pea08,Gal11,Pea12,Gui18b,Cas19} We considered both the planar and twisted (see below) $C_{2v}$ structures here, as in Tozer's 2012 work. \cite{Pea12} 
In the former configuration, the two lowest ESs have a local character, whereas the third (2 $B_2$) and fourth (3 $A_1$) transitions have strong CT characters, with a $r^{\text{eh}}_{\text{ADC}}$ as large as $3.5$ \AA\ for 
the latter. With the {\DZ} basis set, the CCSDT energies are bracketed by the CC3 and CCSDT-3 values that are slightly too small and too large respectively. Our TBEs are $5.53$ and $6.04$ eV with {\TZ}, and \hl{5.32 and 5.86} eV
with {\AVQZ}. The former are close to the extrapolated CCSDR(3)/{\TZ} values of Ref.~\citenum{Cas19}, whereas the latter are significantly larger than the extrapolated CC3/{\AVTZ} estimates of $5.21$ and $5.69$ eV given
in Ref.~\citenum{Gui18b}.

\subsubsection{Phthalazine}

In this asymmetric bicyclic system, the two lowest ESs, of $n \rightarrow \pi^\star$ character, do involve a moderate CT character according to the selected metrics. The data listed in Table \ref{Table-1} show that
CC3 and CCSDT do agree very well, whereas the CCSDT-3 transition energies seem too large by approximately $0.1$ eV. Our TBE/{\TZ} of $3.93$ and $4.34$ eV are most probably trustworthy for this basis set and decrease only very slightly 
with the addition of diffuse basis functions. The most refined previous estimates we are aware of are the CC2/{\AVDZ} results of  Etinski and Marian \cite{Eti17} and the CASPT2/{\QZ} values of Mori and 
coworkers,\cite{Mor09} and it seems reasonable to state that the TBEs gathered in Table \ref{Table-1} are more accurate. Our vertical energies are, as expected, larger than both experimental peak positions \cite{Vas78,Kai78} 
and 0-0 energies. \cite{Inn88}

\subsubsection{Quinoxaline}

In quinoxaline, we identified three transitions possessing a partial CT character with the selected basis set and models: two  $\pi \rightarrow \pi^\star$ ESs as well as a higher-lying $n \rightarrow \pi^\star$ ES. Confirming the trends obtained above, one notes that
the CCSDT excitations energies are roughly in between their CCSDT-3 and CC3 counterparts for the states with a significant electron-hole separation ($r^{\text{eh}}_{\text{ADC}} > 1.5$ \AA), but closer to
the CC3 results for the transitions with milder CT character ($r^{\text{eh}}_{\text{ADC}} \sim 1.0$ \AA). Unexpectedly, the basis set effects seem significantly larger than for phthalazine. For the lowest transition 
considered, $1B_2$, the present TBE/{\AVQZ} of \hl{4.63} eV exceeds significantly the previous CC2/{\AVDZ} \cite{Eti17}  and CASPT2/{\QZ} \cite{Mor09} results. Finally, for the two ESs for which experimental values 
have been reported, \cite{Kai78,Gla70} the correct positive difference is once more obtained.

\subsubsection{Twisted DMABN and PP}

Finally, we consider DMABN and PP in their twisted conformation in which the orthogonality between the NMe$_2$ or pyrrole group and the phenyl moiety was enforced. The GS structures were optimized
in the $C_{2v}$ symmetry. In such conformation, the donor and acceptor units are effectively electronically uncoupled, and one creates two (DMABN) or four (PP) low-lying CT transitions from the nitrogen lone pair (DMABN) or the pyrrole
$\pi$ system (PP) towards the two lowest phenyl $\pi^\star$ orbitals. This also very crudely  mimics the possible TICT behavior of these compounds. At the {\DZ} level, the CCSDT value falls systematically between the
CC3 and CCSDT-3 results, the respective average errors of these two methods being $-0.04$ eV and $+0.07$ eV for the six ESs computed on the twisted molecules.  In all cases, the basis set effects are rather
limited, the {\TZ} results being only decreased by ca.~$-0.10$ eV when going to {\AVQZ}, \hl{except for the} $A_1$ \hl{transition of PP for which the basis set effects are slightly larger}.  For the twisted DMABN our TBE/{\AVQZ} 
values are slightly below the extrapolated CCSD/{\AVTZ} data obtained by Tozer. \cite{Pea12}  For twisted PP, the same observation holds and the present TBEs are larger than CASPT2/DZP values obtained two decades 
ago.\cite{Pro00} In both cases, direct comparisons with experiment, e.g., fluorescence from the TICT structure, remains beyond reach as GS geometries are considered here rather than the ES geometries.

\subsection{Benchmarks}

\subsubsection{Wavefunction and BSE}

\begin{figure*}
\centering
 \includegraphics[width=\linewidth]{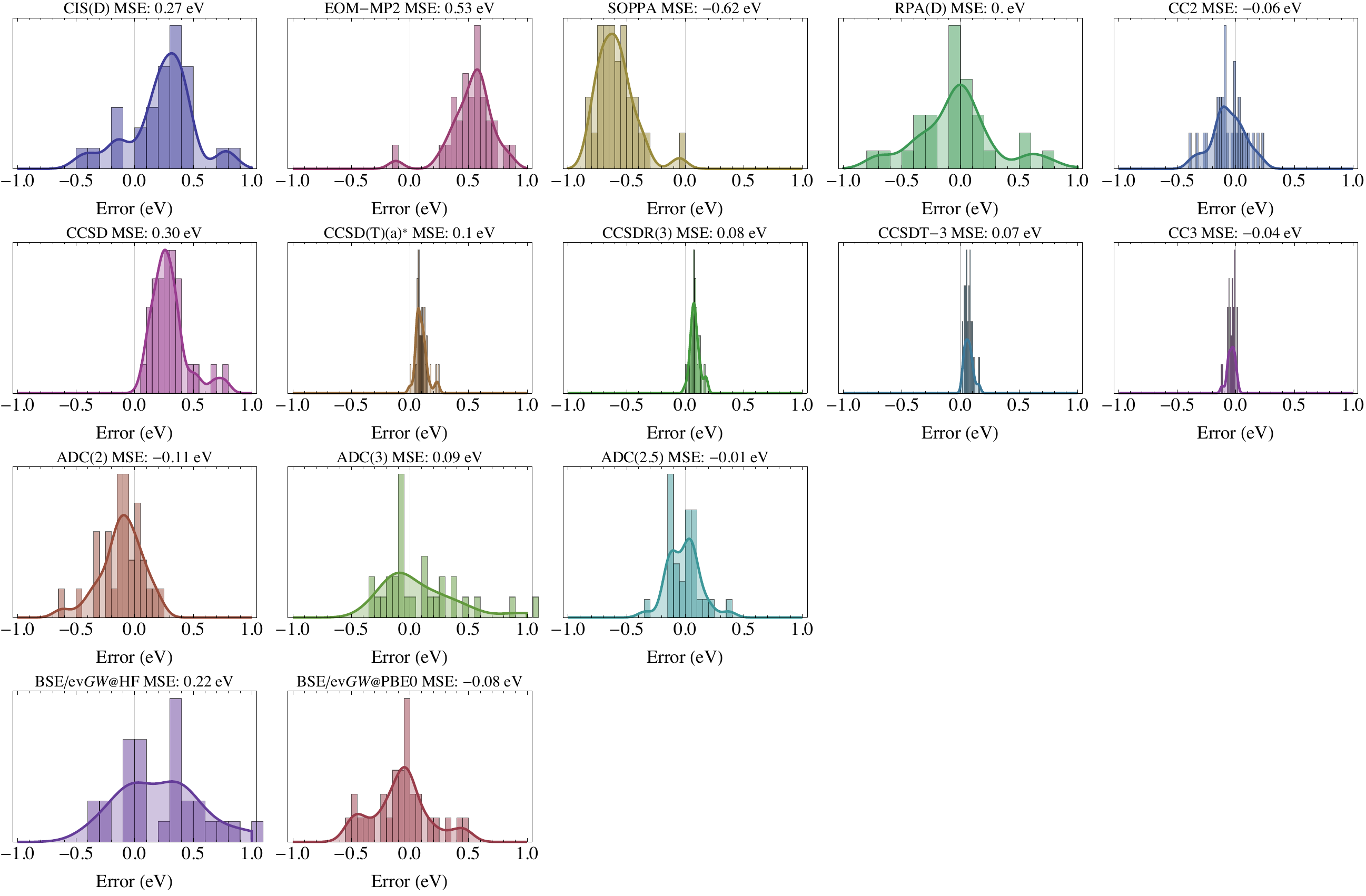}
  \caption{Error patterns against TBE/{\TZ} for wavefunction and BSE approaches.}
   \label{Fig-2}
\end{figure*}

Having a series TBEs of CCSDT quality at hand, it seems natural to investigate the performances of lower-order approaches. Then, we evaluate here wavefunction-, Green's function-, and density-based methods.  For the two former families we rely on the
TBE/{\TZ} data as the basis set dependency is similar for these groups of methods. The correspoding results are collected in Table \ref{Table-2}. For the vast majority of the cases, the identification of the states was 
straightforward for all tested methods, except again for the two peptide derivatives for which a careful inspection of the orbitals/densities was required to reach the correct attribution.  At the bottom of Table \ref{Table-2}, we also
provide statistical quantities obtained by considering these TBEs as reference. We report mean signed error (MSE), mean absolute error (MAE), standard deviation of the errors (SDE), root-mean-square error (RMSE), and maximal positive 
[Max(+)] and negative [Max($-$))] errors. For one transition, our TBE is of CCSDT-3 rather than CCSDT quality, so that the corresponding CCSDT-3 and CC3 results are obviously not included in the statistical analysis (see the footnote in 
Table \ref{Table-2}). Finally, a graphical representation of the error patterns can be found in Fig.~\ref{Fig-2}.

%%% TABLE 2 %%%
\begin{sidewaystable*}
\caption{CT excitation energies (in eV) obtained with various wavefunction-based methods with the cc-pVTZ basis set. 
Statistical quantities are reported at the bottom of the Table. 
For the MSE and MAE, we also provide values obtained for the ``strong CT'' subgroup, i.e., transitions for which $r^{\text{eh}}_{\text{ADC}} \geq 1.75$ \AA.} 
\label{Table-2}
\footnotesize
\vspace{-0.3 cm}
\begin{tabular}{p{3.5cm}l|c|ccccccccccccccc}
\hline 
Molecule 	& State								&TBE	&CIS(D)	&EOM-MP2&\hl{SOPPA}&\hl{RPA(D)}	&CC2	&CCSD	&CCSD(T)(a)*	&CCSDR(3)	&CCSDT-3	&CC3	&ADC(2)	&ADC(3)	&ADC(2.5)	& BSE@HF	& BSE@PBE0	\\
\hline
Aminobenzonitrile&$ 2 A_1$ ($\pi \rightarrow \pi^\star$)	&5.26	&5.57	&5.61	&\hl{4.62}	&\hl{5.21}	&5.26	&5.41	&5.32		&5.31		&5.30		&5.25	&5.16	&5.09	&5.12		&5.22	&5.11	\\
Aniline	&2 $A_1$ ($\pi \rightarrow \pi^\star$)			&5.87	&6.19	&6.12	&\hl{5.27}	&\hl{5.85} &5.86	&5.99	&5.91		&5.90		&5.90		&5.86	&5.79	&5.74	&5.76		&5.80	&5.64	\\
Azulene	&2 $A_1$ ($\pi \rightarrow \pi^\star$)			&3.89	&4.14	&4.37	&\hl{3.27}	&\hl{3.93}	&3.94	&4.02	&3.98		&3.98		&3.92		&3.88	&3.86	&3.65	&3.75		&3.57	&3.45	\\
		&2 $B_2$ ($\pi \rightarrow \pi^\star$)			&4.55	&4.78	&5.23	&\hl{4.05}	&\hl{5.14}	&4.69	&4.82	&4.69		&4.68		&4.61		&4.52	&4.67	&4.45	&4.56		&4.64	&4.40	\\
Benzonitrile&1 $A_2$ ($\pi_{\text{CN}} \rightarrow \pi^\star$)
											&7.10	&7.85	&7.52	&\hl{6.76}	&\hl{7.69}	&7.32	&7.33	&7.17 		&7.16		&7.15		&7.08	&7.28	&6.77	&7.03		&7.03	&6.58	\\		
Benzothiadiazole&1 $B_2$ ($\pi \rightarrow \pi^\star$)	&4.37	&4.74	&5.03	&\hl{3.79}	&\hl{4.37}	&4.47	&4.63	&4.45 		&4.44		&4.40		&4.30	&4.46	&4.04	&4.25		&4.16	&3.89	\\
Dimethylaminobenzonitrile	
		&2 $A_1$ ($\pi \rightarrow \pi^\star$)			&4.94	&5.26	&5.33	&\hl{4.20} 	&\hl{4.92}	&4.85	&5.10	&5.00		&4.99		&4.99		&4.93	&4.73	&4.87	&4.80		&4.97	&4.89	\\
Dimethylaniline&1 $B_2$ ($\pi \rightarrow \pi^\star$)		&4.47	&4.67	&4.90	&\hl{3.92}	&\hl{4.22}	&4.49	&4.66	&4.54		&4.55		&4.53		&4.48	&4.47	&4.51	&4.49		&4.78	&4.56	\\
			&2 $A_1$ ($\pi \rightarrow \pi^\star$)		&5.54	&5.95	&5.85	&\hl{4.85}	&\hl{5.63}	&5.44	&5.68	&5.59		&5.58		&5.58		&5.53	&5.35	&5.52	&5.44		&5.58	&5.42	\\
Dipeptide	& 7 $A''$	($n_1 \rightarrow \pi_2^\star$)		&8.15	&9.51	&9.01	&\hl{7.48}	&\hl{9.37}	&7.89	&8.92	&8.37		&8.33		&8.31		&8.04	&7.82	&9.22	&8.52		&9.03	&8.59	\\%33-39/ ou 28A'=>8A"	
$\beta$-Dipetide&7 $A'$ ($\pi_1 \rightarrow \pi_2^\star$)	&8.51	&8.60	&9.00	&\hl{7.96}	&\hl{8.44}	&8.34	&8.90	&8.63		&8.59$^a$	&8.59		&8.46$^b$&8.30	&8.88	&8.59		&9.12	&8.84	\\%39-43 (0.494); 38-43 (-0.241)	<= CCSD combo/TZ  13 -- 10.787	Dalton	8eme A" => 1 A"%39=8A" (n¡8 de sym2)=> 43=9A" (=n¡1 de sym 2)
		&10 $A''$($n_1 \rightarrow \pi_2^\star$)		&8.90	&9.71	&9.63	&\hl{8.16}	&\hl{9.67}	&8.52	&9.58	&9.14		&9.08$^a$	&9.02		&8.78$^b$&8.45	&9.76$^c$&9.10		&9.67	&9.38	\\%37-43 (-0.587); 37-46 (0.215)	<= CCSD combo/TZ 11.8578%37=31A' (n¡21 de sym1)=> 43=9A" (=n¡1 de sym 2)
Hydrogen Chloride&1 $\Pi$ ($n  \rightarrow \sigma^\star$)	&8.10	&8.32	&7.98	&\hl{8.05}	&\hl{8.07}	&8.28	&8.18	&8.10 		&8.10		&8.12		&8.11	&8.30	&8.02	&8.16		&8.41	&7.72	\\
Nitroaniline	& 2 $A_1$ ($\pi \rightarrow \pi^\star$)	&4.57	&4.75	&5.15	&\hl{3.85}	&\hl{4.36}	&4.55	&4.80	&4.67 		&4.65		&4.61		&4.51	&4.44	&4.42	&4.43		&4.55	&4.47	\\		
Nitrobenzene& 2 $A_1$ ($\pi \rightarrow \pi^\star$)		&5.57	&5.93	&6.11	&\hl{4.95}	&\hl{5.57}	&5.63	&5.77	&5.67 		&5.64		&5.59		&5.52	&5.55	&5.31	&5.43		&5.47	&5.25	\\	
Nitrodimethylaniline&2 $A_1$ ($\pi \rightarrow \pi^\star$)	&4.28	&4.44	&4.89	&\hl{3.47}	&\hl{4.06}	&4.18	&4.53	&4.39 		&4.37		&4.33		&4.22	&4.05	&4.21	&4.13		&4.30	&4.28	\\
Nitropyridine N-Oxide&2 $A_1$ ($\pi \rightarrow \pi^\star$)&4.24	&4.26	&4.71	&\hl{2.88}	&\hl{4.52}	&4.10	&4.46	&4.31 		&4.28		&4.24$^d$	&4.13$^d$&3.62	&4.17	&3.90		&3.85	&4.05	\\	
N-Phenylpyrrole&2 $B_2$  ($\pi \rightarrow \pi^\star$)	&5.53	&5.98	&6.06	&\hl{5.08}	&\hl{5.62}	&5.55	&5.84	&5.62 		&5.61		&5.60		&5.50	&5.57	&5.46	&5.52		&5.59	&5.29	\\
			&3 $A_1$ ($\pi \rightarrow \pi^\star$)		&6.04	&6.35	&6.76	&\hl{5.66}	&\hl{6.23}	&6.02	&6.52	&6.18 		&6.16		&6.14		&5.97	&6.07	&6.16	&6.11		&6.35	&6.03	\\
Phthalazine	&1 $A_2$ ($n \rightarrow \pi^\star$)		&3.93	&4.31	&4.47	&\hl{3.24}	&\hl{3.91}	&3.78	&4.26	&4.05 		&4.04		&4.03		&3.89	&3.79	&4.19	&3.99		&4.39	&3.92	\\
			&1 $B_1$ ($n \rightarrow \pi^\star$)		&4.34	&4.75	&4.93	&\hl{3.61}	&\hl{4.27}	&4.22	&4.64	&4.46 		&4.46		&4.43		&4.32	&4.23	&4.49	&4.36		&4.73	&4.28	\\
Quinoxaline	&1 $B_2$ ($\pi \rightarrow \pi^\star$)		&4.74	&5.17	&5.32	&\hl{4.05}	&\hl{4.92}	&4.66	&5.00	&4.82 		&4.81		&4.79		&4.69	&4.65	&4.64	&4.64		&4.53	&4.28	\\
			&3  $A_1$ ($\pi \rightarrow \pi^\star$)	&5.75	&5.90	&6.36	&\hl{5.23}	&\hl{5.43}	&5.83	&6.01	&5.86 		&5.86		&5.84		&5.76	&5.82	&5.59	&5.71		&6.83	&5.70	\\
			&2 $B_1$ ($n \rightarrow \pi^\star$)		&6.33	&6.81	&7.13	&\hl{5.79}	&\hl{6.45}	&6.17	&6.87	&6.50 		&6.49		&6.48		&6.26	&6.25	&6.79	&6.52		&6.88	&6.42	\\
Twisted DMABN&1 $A_2$($n \rightarrow \pi^\star$)		&4.17	&3.82	&4.55	&\hl{3.36}	&\hl{3.49}	&3.89	&4.41	&4.23 		&4.23		&4.25		&4.15	&3.84	&4.56	&4.20		&4.61	&4.32	\\
			&1 $B_1$($n \rightarrow \pi^\star$)		&4.84	&4.38	&5.33	&\hl{4.07}	&\hl{4.12}	&4.50	&5.19	&4.91 		&4.91		&4.95		&4.81	&4.50	&5.40	&4.95		&5.34	&5.05	\\
Twisted PP	&2 $B_2$  ($\pi \rightarrow \pi^\star$)	&5.73	&5.55	&6.37	&\hl{5.23}	&\hl{5.35}	&5.64	&6.10	&5.79 		&5.79		&5.80		&5.67	&5.67	&5.87	&5.77		&6.04	&5.65	\\
			&2 $A_1$  ($\pi \rightarrow \pi^\star$)	&5.82	&6.09	&6.41	&\hl{5.41}	&\hl{5.81}	&5.78	&6.18	&5.92 		&5.91		&5.89		&5.76	&5.82	&5.93	&5.87		&6.07	&5.80	\\
			&1 $A_2$  ($\pi \rightarrow \pi^\star$)	&6.04	&5.90	&6.63	&\hl{5.43}	&\hl{5.56}	&5.92	&6.35	&6.11 		&6.11		&6.12		&6.01	&5.90	&6.24	&6.07		&6.41	&6.04	\\
			&1 $B_1$  ($\pi \rightarrow \pi^\star$)	&6.28	&6.16	&6.86	&\hl{5.69}	&\hl{5.91}	&6.14	&6.61	&6.41 		&6.38		&6.36		&6.24	&6.13	&6.57	&6.35		&6.60	&6.27	\\
\hline	
MSE			&								&		&0.27	&0.53	&\hl{-0.62}	&\hl{ 0.01}&-0.06	&0.30	&0.10		&0.08		&0.07		&-0.04	&-0.11	&0.09	&-0.01		&0.22	&-0.08\\
MSE (strong CT)				&				&		&0.23	&0.60	&\hl{-0.67}	&\hl{-0.02}	&-0.14	&0.37	&0.12		&0.10		&0.09		&-0.05	&-0.19	&0.25	&0.03		&0.31	&0.03\\
MAE			&								&		&0.35	&0.53	&\hl{0.62}	&\hl{0.27}	&0.12	&0.30	&0.10		&0.08		&0.07		&0.04	&0.16	&0.25	&0.11		&0.32	&0.20\\
MAE (strong CT)				&				&		&0.37	&0.60	&\hl{0.67}	&\hl{0.34}	&0.15	&0.37	&0.12		&0.10		&0.09		&0.05	&0.19	&0.30	&0.11		&0.38	&0.16\\
SDE			&								&		&0.35	&0.19	&\hl{0.21}	&\hl{0.40}	&0.14	&0.16	&0.05		&0.04		&0.04		&0.03	&0.18	&0.33	&0.14		&0.35	&0.25\\
RMSE		&								&		&0.43	&0.56	&\hl{0.65}	&\hl{0.40}	&0.15	&0.33	&0.11		&0.09		&0.08		&0.05	&0.21	&0.34	&0.13		&0.41	&0.26\\
Max($+$)		&								&		&1.36	&0.86	&\hl{-0.05}	&\hl{1.22}	&0.22	&0.77	&0.24		&0.18		&0.16		&0.01	&0.20	&1.07	&0.37		&1.08	&0.48\\
Max($-$)		&								&		&-0.46	&-0.12	&\hl{-1.36}	&\hl{-0.72}	&-0.38	&0.08	&0.00		&0.00		&0.02		&-0.12	&-0.62	&-0.33	&-0.34		&-0.39	&-0.52\\
\hline																	
\end{tabular}
\vspace{-0.3 cm}
\begin{flushleft}
\begin{scriptsize}
$^a${Basis set extrapolated CCSDR(3)/{\TZ} values from Ref. \citenum{Cas19};}
$^b${CC3/{\DZ} value corrected by the difference between CCSDT-3/{\TZ} and CCSDT-3/{\DZ} values;}
$^c${ADC(3)/{\DZ} value corrected by the difference betweenADC(2)/{\TZ} and ADC(2)/{\DZ} values;}
$^d${Not included in the benchmark statistics.}
\end{scriptsize}
\end{flushleft}
\end{sidewaystable*}
%%% %%% %%% %%%

As can be seen in Table \ref{Table-2} and Fig.~\ref{Fig-2}, CIS(D) typically overestimates transition energies, except for the twisted compounds. Hence, this leads to a quite large MSE value of $0.27$ eV. The CIS(D) MAE, $0.35$ eV,  significantly exceeds the $0.22$ eV value
reported for the local transitions of the very large QUEST database, \cite{Ver21} hinting that CT states are likely difficult for CIS(D).  \hl{RPA(D) is somehow superior as its MSE is close to zero, and its MAE is smaller, 0.34 eV, though the differences seem to decrease
when the CT character increase. A similar MAE of 0.35 eV was reported with RPA(D) for the valence singlet transitions of Thiel's set.}\cite{Sau15} EOM-MP2, another ``computationally light'' method also named CCSD(2) in some works, systematically overshoots the transition 
energy (except for HCl), the MAE being very large ($> 0.50$ eV). Note, however, that a quite systematic error pattern is obtained (see Fig.~\ref{Fig-2}), as shown by the very acceptable SDE of $0.19$ eV. This typical feature of EOM-MP2 (clear overestimation of the transition 
energies with a significant increase of the magnitude of the error with system size) was also clearly identified in our previous benchmarks focussing on Rydberg and local transitions. \cite{Ver21}  For their intermolecular CT set, Kozma and coworkers reported a very similar 
SDE ($0.15$ eV), but a smaller MSE ($0.31$ eV) for the same method, a clear overestimation trend being also found. \cite{Koz20} \hl{SOPPA mirrors somehow the behavior of EOM-MP2, with strong underestimations (MSE of $-0.62$ eV), but in a rather systematic way, 
so that the SDE is also rather small ($0.21$ eV). We note that the fact the SOPPA underestimates transition energies was already reported in several benchmarks,} \cite{Hed13,Sau15,Haa19} \hl{and is not specific to CT states, though the errors are particularly large here.}

As expected,  \cite{Har14} CC2 and ADC(2) excitation energies are highly correlated (the $R^2$ between the two series of transition 
energies attain $0.995$), and we found that the former method has a slight edge in terms of accuracy. For the present CT set, CC2 and ADC(2) are also more accurate than both CIS(D) and EOM-MP2, with MAE of $0.12$ eV (CC2) and $0.16$ eV [ADC(2)].  Various research
groups reported similar average errors for local transitions in molecules of similar sizes. \cite{Sch08,Sil10c,Win13,Har14,Jac15b,Kan17,Loo20a,Loo20b,Ver21}  Interestingly, when considering only the subset of strong CT ($r^{\text{eh}}_{\text{ADC}} \geq 1.75$ \AA), 
one notes larger errors with significant (and nearly systematic) underestimations leading to negative MSEs of $-0.14$ eV and $-0.19$ eV for CC2 and ADC(2), respectively. In other words, both methods tend to undershoot the CT transition energies when the electron-hole separation becomes
sizable. This trend is fully consistent with the investigation of Kozma and coworkers devoted to  intermolecular CT ESs: \cite{Koz20} they reported MSE of $-0.36$ eV for both methods.  

The contrast is clear with CCSD that overestimates quite considerably the transition energies, especially for the strong CT subset with a MSE of $+0.37$ eV. On the brighter side, CCSD provides quite systematic errors with a SDE of $0.16$ eV. These trends are typical of CCSD and were reported in several 
benchmarks considering local and Rydberg excitations. \cite{Sch08,Car10,Wat13,Kan14,Jac17b,Kan17,Dut18,Jac18a,Loo18a,Loo20a,Loo20b,Ver21} A MSE of $+0.30$ eV and a SDE of $0.08$ eV have been reported for the 14 intermolecular CT transitions 
of Ref.~\citenum{Koz20}. 

The results obtained with the three CC methods including contributions from the triples are much more satisfying. Indeed, the MAEs are of the order of $0.10$ eV (or smaller) and the SDEs are below the $0.05$ eV threshold. A near-perfect correlation between
the CCSD(T)(a)*, CCSDR(3), and CCSDT-3 transitions is noticeable ($R^2$ larger than $0.999$ for all possible pairs of methods, see also Fig.~\ref{Fig-2}). Comparing the two approaches with perturbative triples, namely, CCSD(T)(a)* and CCSDR(3), 
one notes very similar deviations, with a slight edge for the second method.  Consistently with the results discussed above, CCSDT-3 systematically overestimates the TBEs, whereas CC3 tends to provide slightly too small values. While the sign and magnitude of the 
differences between the excitation energies obtained with these two CC approaches nicely parallel the findings of Kozma \emph{et al.}, \cite{Koz20} we find, in contrast to their work, that CC3 is superior as it provides chemically accurate CT excitation energies.
This statement holds also when considering only the strong CT subset, and is in line with the results obtained for local and Rydberg transitions of compounds of similar size. \cite{Loo20a,Ver21}  

Consistently with our recent investigations, \cite{Loo20b,Ver21} ADC(3) does not significantly improve over ADC(2), as it yields large overestimations for the strong CT subset with a  MSE of $+0.25$ eV, and a MAE of $+0.30$ eV, together with a significant dispersion (see Fig.~\ref{Fig-2}). Therefore
the trends obtained with ADC(3) are opposite to the ones noticed above for ADC(2). The ADC(2.5) approach \cite{Loo20b} --- which simply consists in taking the average between the ADC(2) and ADC(3) transition energies --- is more accurate than the two other ADC methods with 
a negligible MSE, a MAE of ca.~$0.11$ eV, and a SDE of $0.14$ eV. These values indicate that ADC(2.5) outperforms all the wavefunction methods tested here sharing the same $\mathcal{O}(N^6)$ or a lower $\mathcal{O}(N^5)$ computational scaling. 
These statistical values are totally similar to their local and Rydberg counterparts obtained in the QUEST database (respective MAEs of $0.08$ and $0.09$ eV), \cite{Ver21} hinting that ADC(2.5) might be a valuable compromise for many families of transitions.

%%% FIGURE 2 %%%
%%% %%% %%% %%%
 
%%% TABLE 3 %%% 
\begin{table*}
\caption{TD-DFT transition energies (in eV) obtained with the {\AVQZ} basis set. Statistical quantities are reported at the bottom of the Table. 
See caption of Table of \ref{Table-2} for more details.} 
\label{Table-3}
\footnotesize
\vspace{-0.3 cm}
\begin{tabular}{p{4cm}l|c|cccccccc}
\hline 
Molecule 	& State									&TBE	&B3LYP	&PBE0	&M06-2X	&CAM-B3LYP	
																						&LC-$\omega$HPBE	
																								& $\omega$B97X 	&$\omega$B97X-D	& M11	\\
\hline
Aminobenzonitrile&2 $A_1$ ($\pi \rightarrow \pi^\star$)		&\hl{5.09}	&4.87	&4.96	&5.11	&5.06	&5.22	&5.17	&5.10	&5.11	\\%Faible CT dans la rŽalitŽ en AQZ, surtout en M11...
Aniline	&2 $A_1$ ($\pi \rightarrow \pi^\star$)				&\hl{5.48}	&5.24	&5.37	&5.46	&5.42	&5.62	&5.57	&5.49	&5.28	\\%La plupart des XCFs le donne comme local lˆ...
Azulene	&2 $A_1$ ($\pi \rightarrow \pi^\star$)				&\hl{3.84}	&3.60	&3.67	&3.83	&3.72	&3.82	&3.77	&3.72	&3.83	\\
		&2 $B_2$ ($\pi \rightarrow \pi^\star$)				&\hl{4.49}	&4.62	&4.71	&4.81	&4.76	&4.85	&4.82	&4.76	&4.84	\\%essayer une doublue hybrid pour le fun ? TTe les XCF frappe trop haut...
Benzonitrile&1 $A_2$ ($\pi_{\text{CN}} \rightarrow \pi^\star$)	&\hl{7.05}	&6.20	&6.29	&6.24	&6.59	&6.85	&6.77	&6.60	&6.75	\\
Benzothiadiazole&1 $B_2$ ($\pi \rightarrow \pi^\star$)		&\hl{4.28}	&3.78	&3.89	&4.19	&4.10	&4.39	&4.28	&4.13	&4.34	\\
Dimethylaminobenzonitrile&2 $A_1$ ($\pi \rightarrow \pi^\star$)	&\hl{4.86}	&4.65	&4.74	&4.93	&4.90	&5.07	&5.02	&4.93	&4.99	\\
Dimethylaniline&1 $B_2$ ($\pi \rightarrow \pi^\star$)			&\hl{4.40}	&4.41	&4.50	&4.71	&4.67	&4.82	&4.77	&4.67	&4.75	\\
			&2 $A_1$ ($\pi \rightarrow \pi^\star$)			&\hl{5.40}	&5.22	&5.31	&5.46	&5.42	&5.57	&5.53	&5.45	&5.47	\\
Hydrogen Chloride&1 $\Pi$ ($n  \rightarrow \sigma^\star$)		&\hl{7.88}	&7.32	&7.57	&7.54	&7.50	&7.94	&7.86	&7.68	&7.34	 \\
Nitroaniline	& 2 $A_1$ ($\pi \rightarrow \pi^\star$)		&\hl{4.39}	&3.92	&4.08	&4.45	&4.34	&4.68	&4.59	&4.41	&4.59	\\
Nitrobenzene& 2 $A_1$ ($\pi \rightarrow \pi^\star$)			&\hl{5.39}	&4.72	&4.89	&5.30	&5.11	&5.46	&5.34	&5.17	&5.39	\\
Nitrodimethylaniline&2 $A_1$ ($\pi \rightarrow \pi^\star$)		&\hl{4.13}	&3.67	&3.82	&4.23	&4.14	&4.49	&4.41	&4.22	&4.38	\\
Nitropyridine N-Oxide&2 $A_1$ ($\pi \rightarrow \pi^\star$)	&\hl{4.10}	&3.81	&3.95	&4.22	&4.14	&4.34	&4.30	&4.22	&4.33	\\
N-Phenylpyrrole&2 $B_2$  ($\pi \rightarrow \pi^\star$)		&\hl{5.32}	&4.73	&4.86	&5.24	&5.22	&5.66	&5.52	&5.28	&5.44	\\
			&3 $A_1$ ($\pi \rightarrow \pi^\star$)			&\hl{5.86}	&4.92	&5.09	&5.83	&5.90	&6.81	&6.52	&6.02	&6.45	\\
Phthalazine	&1 $A_2$ ($n \rightarrow \pi^\star$)			&\hl{3.91}	&3.52	&3.65	&4.02	&4.05	&4.27	&4.25	&4.03	&4.08	\\
			&1 $B_1$ ($n \rightarrow \pi^\star$)			&\hl{4.31}	&3.94	&4.05	&4.24	&4.39	&4.59	&4.57	&4.38	&4.32	\\
Quinoxaline	&1 $B_2$ ($\pi \rightarrow \pi^\star$)			&\hl{4.63}	&4.08	&4.20	&4.61	&4.50	&4.87	&4.73	&4.53	&4.77	\\
			&3  $A_1$ ($\pi \rightarrow \pi^\star$)		&\hl{5.65}	&5.70	&5.81	&5.96	&5.91	&6.08	&6.02	&5.93	&6.03	\\
			&2 $B_1$ ($n \rightarrow \pi^\star$)			&\hl{6.22}	&5.69	&5.86	&6.44	&6.46	&6.94	&6.80	&6.44	&6.50	\\
Twisted DMABN&1 $A_2$($n \rightarrow \pi^\star$)			&\hl{4.12}	&3.20	&3.34	&3.96	&3.95	&4.38	&4.28	&3.96	&4.11	\\
			&1 $B_1$($n \rightarrow \pi^\star$)			&\hl{4.75}	&3.87	&4.04	&4.81	&4.72	&5.38	&5.14	&4.69	&5.03	\\
Twisted PP	&2 $B_2$  ($\pi \rightarrow \pi^\star$)		&\hl{5.58}	&4.34	&4.54	&5.29	&5.33	&6.32	&6.05	&5.40	&5.96	\\%WBX-D strong mix with other B2 state at 5.59....faire un warning somewhere...
			&2 $A_1$  ($\pi \rightarrow \pi^\star$)		&\hl{5.65}	&4.43	&4.64	&5.47 	&5.54	&6.52 	&6.10	&5.65	&6.09	\\%LC-wHPBE= mix trs fort, pas facile, quasi impossible...
			&1 $A_2$  ($\pi \rightarrow \pi^\star$)		&\hl{5.95}	&4.97	&5.17	&5.81	&5.90	&6.56	&6.40	&6.00	&5.78	\\
			&1 $B_1$  ($\pi \rightarrow \pi^\star$)		&\hl{6.17}	&5.08	&5.31	&6.08	&6.14	&6.90	&6.70	&6.25	&		\\
\hline		
MSE			&									&		&-0.5\hl{3}	&-0.39	&-0.0\hl{2}	&-0.04	&0.35	&0.2\hl{4}	&0.01	&0.12	\\
MSE (strong CT)	&								&		&-0.73	&-0.57	&-0.03	&-0.0\hl{2}	&0.5\hl{1}	&0.35	&0.0\hl{3}	&0.2\hl{1}	\\
MAE			&									&		&0.55	&0.43	&0.1\hl{5}	&0.1\hl{4}	&0.37	&0.27	&0.1\hl{3}	&0.22	\\
MAE (strong CT)		&							&		&0.73	&0.57	&0.1\hl{2}	&0.1\hl{0}	&0.5\hl{1}	&0.35	&0.1\hl{0}	&0.23	\\
SDE			&									&		&0.3\hl{8}	&0.3\hl{5}	&0.23	&0.1\hl{8}	&0.2\hl{8}	&0.22	&0.1\hl{7}	&0.2\hl{5}	\\
RMSE		&									&		&0.6\hl{5}	&0.5\hl{2}	&0.2\hl{2}	&0.19	&0.4\hl{5}	&0.32	&0.17	&0.27	\\
Max($+$)		&									&		&0.1\hl{3}	&0.2\hl{2}	&0.3\hl{2}	&0.2\hl{7}	&0.9\hl{5}	&0.6\hl{6}	&0.2\hl{8}	&0.5\hl{9}		\\
Max($-$)		&									&		&-1.2\hl{4}	&-1.0\hl{4}	&-0.8\hl{1}	&-0.4\hl{6}	&-0.\hl{20}	&-0.2\hl{8}	&-0.4\hl{5}	&-0.54	\\
\hline
\end{tabular}
\vspace{-0.3 cm}
\begin{flushleft}
\begin{scriptsize}
\end{scriptsize}
\end{flushleft}
\end{table*}
%%% %%% %%% %%%

The BSE/ev$GW$ calculations were performed with two very different sets of eigenstates (HF and PBE0) as input. The well-known positive impact of the ev$GW$ procedure\cite{Jac15a,Jac15b,Jac17b,Gui18b}  undoubtedly emerges in  Table \ref{Table-2}.
Indeed, one obtains a mean absolute deviation (MAD) of $0.33$ eV only between the two sets whereas much larger variations would be reached by comparing TD-PBE0 and TD-HF. There is also a strong correlation ($R^2$ of $0.981$) between the 
two sets of transition energies. In terms of performances, it is clearly and unsurprisingly more favorable to perform the ev$GW$ calculations on the basis of KS orbitals. Indeed, for the full set, BSE/ev$GW$@PBE0 yields a MAE of $0.20$ eV 
($0.16$ eV for the strong CT subset), which is somewhat similar to the  ADC(2) performances, though BSE/ev$GW$@PBE0 yields less consistent results than these two wavefunction theories with a non-negligible SDE of $0.28$ eV. One should recall here 
that BSE/ev$GW$@PBE0 scales as $\mathcal{O}(N^4)$, and can be applied to very large systems, so that its global accuracy for the CT transition remains very satisfying in comparison to the associated computational cost.
While TD-DFT calculations (see next Subsection)  can certainly offer a similar accuracy with proper tuning (range-separation parameter, amount of short/long-range exact exchange) for this specific family of systems, the BSE formalism which relies on the electron-hole 
screened Coulomb potential instead of the exchange-correlation kernel, allows to tackle with similar accuracy local, \cite{Jac15a,Bru15,Gui18b,Ngu19,Chi20} cyanines, \cite{Bou14} and CT excitations. This is an important property in the present case of intramolecular 
CT excited states that show weak to strong (Frenkel) character. Finally, we note that the quality of the BSE transition energies is also closely related to the ones of the quasiparticles. \cite{Bru15,Gui18b} This also holds at the TD-DFT level. \cite{Shu20}

\subsubsection{TD-DFT}
	 
\begin{figure}
\centering
 \includegraphics[width=\linewidth]{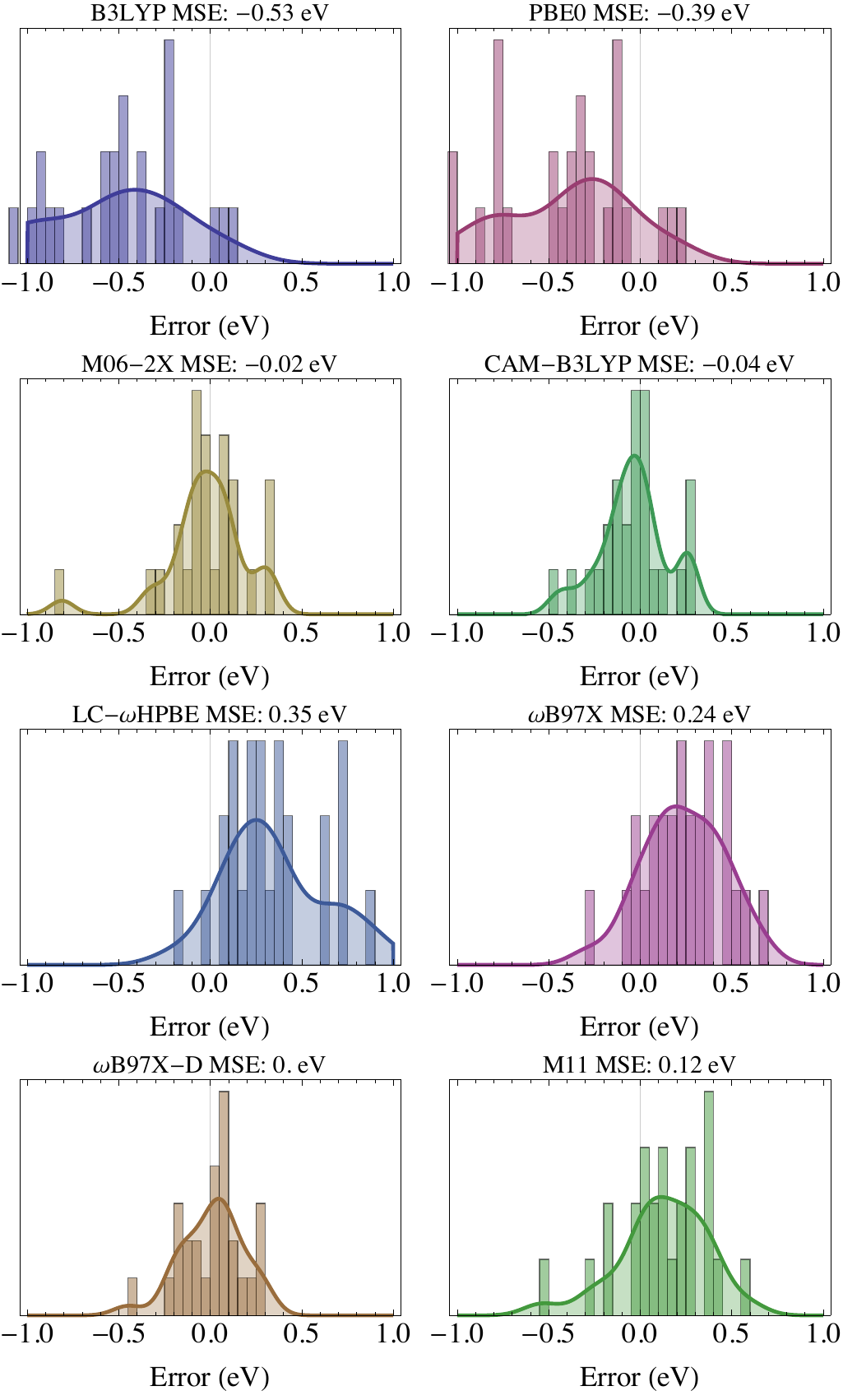}
  \caption{Error patterns against TBE/{\AVQZ} for TD-DFT relying on several XCFs.}
   \label{Fig-3}
\end{figure}

Let us now turn towards the TD-DFT results listed in Table \ref{Table-3} and displayed in Fig.~\ref{Fig-3}. In this case, we switch to the TBE/{\AVQZ} reference data in order to have a fairer comparison between the density-based TD-DFT method and the wavefunction methods employed to produce these TBEs.
Indeed, it is well known that TD-DFT is less sensitive to 
the basis set size than wavefunction approaches. Therefore, near-CBS limit excitation energies seem to be the only option for a trustworthy comparison between these two families. This choice is however not without consequences. First, TD-DFT calculations become 
unnecessarily expensive. Second, and more importantly, this extended basis set which contains diffuse basis functions yields significant orbital mixing at the TD-DFT level, blurring the precise nature of the ESs and making the attribution of the various 
transitions more challenging, a drawback especially marked with LC-$\omega$HPBE and M11.

%%% FIGURE 3 %%%
%%%  %%% %%% %%%

As expected from previous CT benchmarks, \cite{Toz03,Dre04,Pea08,Pea12,Lau13} B3LYP and PBE0 tend to underestimate vertical transition energies with errors of the order of $-1.0$ eV for the pathological cases characterized by a negligible 
 overlap between the occupied and virtual MOs involved in the transition.  Likewise, the fact that RSHs tend to be more accurate than B3LYP or PBE0 is no surprise for a benchmark study focussing on CT transitions.
Nonetheless, the statistical quantities listed at the bottom of Table \ref{Table-3} show some interesting trends. First, M06-2X, a global hybrid containing 54\%\ of exact exchange, performs well with a MSE of -0.0\hl{2} eV and a 
MAE of 0.1\hl{5} eV (even 0.1\hl{2} eV for the states in which $r^{\text{eh}}_{\text{ADC}} \geq 1.75$ \AA). Although the SDE ($0.23$ eV) and the largest negative deviation (-0.8\hl{1} eV for benzonitrile) are sizable, it appears that M06-2X provides 
quite accurate CT transition energies even for the difficult twisted compounds. Given that the same XCF was shown to be efficient for several other families of transitions, \cite{Jac12b,Lea12,Ise12,Jac15b} M06-2X appears as 
a handy  ``Swiss army knife'' approach in the TD-DFT framework, though it is not suited for CT excitations where the distance between the hole and the electron is very large.  

If one now turns towards the RSH family, one notices that the smallest statistical deviations
are  obtained with $\omega$B97X-D with MSE close to zero, MAE below 0.15 eV, combined  with SDE and RMSE smaller than $0.20$ eV.  CAM-B3LYP exhibits very similar  performances although the MSEs are slightly negative, 
likely due to the use of ``only'' 65\%\ of exact exchange at long-range in CAM-B3LYP instead of 100\%\ in $\omega$B97X-D. This induces slight underestimations of the transition energies for the twisted compounds with 
CAM-B3LYP. The MAE that we determined for CAM-B3LYP (0.1\hl{4} eV) is significantly smaller than the ones reported in both Refs.~\citenum{Pea08} ($0.27$ eV) and \citenum{Cas19}  ($0.46$ eV). This might be explained by the fact that the very challenging di- and tri-peptides 
were included in these previous works.  For our test set, the three other RSHs, which include larger shares of exact exchange tend to be less effective with MAEs of $0.22$ eV (M11), 
0.27 eV ($\omega$B97X), and $0.37$ eV (LC-$\omega$HPBE), and positive MSEs for these three functionals.  It is noteworthy that the MAE reported in Ref.~\citenum{Cas19} for $\omega$B97X is slightly smaller ($0.23$ eV). Of course,
one should keep in mind that the performances of a specific XCF within TD-DFT is very dependent on the nature of the considered transitions. Therefore, the error bars obtained here are likely only relevant for similar intramolecular CT 
excitations.

\section{Conclusions}

We have considered, in a series of $\pi$-conjugated compounds, a set of thirty electronic excitation energies presenting a mild to strong intramolecular CT character, the electron-hole separation induced by the electronic
transition spanning from $0.83$ to $4.35$ \AA\ according to an analysis of the ADC(2) transition densities.  Using ground state geometries determined at the CC3/{\TZ} or CCSD(T)/{\TZ} level, we have defined theoretical best estimates (TBEs) 
for these vertical transition energies by correcting CSDT/{\DZ} values by the difference between CC3/{\TZ} and CC3/{\DZ} [see Eq.~\eqref{eq1}]. These TBEs were further extended to {\AVTZ} and  {\AVQZ} by applying a similar basis set correction approach using \hl{CCSDT-3, CCSD
and} CC2 transition energies [see Eqs.~\eqref{eq2}~and~\eqref{eq3}]. 

For almost every compounds and states considered here, the present TBEs are the most accurate published to date.  Although higher-level calculations (e.g., CCSDTQ) were not technically feasible in the present context, the fact that highly consistent CCSDT-3, CC3, and CCSDT 
values were almost systematically obtained provides strong confidence in the quality of the present reference data. In more details, for excitations of mild CT character, CC3 and CCSDT transition energies are typically highly similar, as for local valence transitions, \cite{Ver21} 
whereas for transitions with more pronounced CT nature, the CCSDT energies are bracketed by CCSDT-3 (high) and CC3 (low). In ``pure'' intermolecular CT excitation, CCSDT-3 was in fact found to better match CCSDT than CC3. \cite{Koz20}

We hope that the present reference energies will be useful for the electronic structure community developing and assessing new ES theories. As a first step in this direction, we have benchmarked ten popular wavefunction methods, the Bethe-Salpeter equation (BSE) formalism from many-body perturbation theory, as well as 
TD-DFT with various global and range-separated hybrid functionals.  The four CC models including contributions for the triple excitations [i.e., CCSD(T)(a)*, CCSDR(3), CCSDT-3, and CC3] deliver very solid results with small errors and highly-consistent  excitation energies.  Amongst these 
approaches, CC3 is the only one providing chemically accurate excitation energies (error below $0.043$ eV). The computational cost of these methods is however high.  Regarding computationally cheaper methods with a formal $\mathcal{O}(N^6)$ scaling with system size, it turns out 
that (EOM-)CCSD  overestimates the transition energies significantly but with rather systematic errors, whereas ADC(2.5) appears to be a valuable alternative for a similar computational cost.  Indeed, ADC(2.5) delivers a MAE of $\sim 0.10$ eV.  
Amongst the $\mathcal{O}(N^5)$ methods, CC2 is the most effective with typical underestimations of roughly $-0.15$ eV but consistent estimates. ADC(2) yields similar, yet slightly less accurate, results than CC2. \hl{Amongst the computationally
efficient wavefunction schemes, RPA(D) seems to be a reasonable choice.} For the most effective  $\mathcal{O}(N^4)$  approaches \hl{allowing calculations on very large systems}, one can likely select BSE/ev$GW$@PBE0 which yields consistent estimates for the strong 
CT transitions, though with  a larger dispersion than CC2 or CCSD. With TD-DFT, the most accurate  transition energies are produced with (in decreasing order of accuracy) $\omega$B97X-D, CAM-B3LYP, and M06-2X, all three models providing typical errors of approximately $0.15$ eV, 
but again with slightly higher dispersion than most wavefunction methods. \hl{The interested reader will find in the SI a Table listing the benchmarked methods, their formal scaling, and the obtained MAEs (Table S7). }

The present complementary set of reference energies for CT excited states is currently being merged into the QUEST database of highly-accurate excitation energies which now gathers more than 500 chemically-accurate transition energies. \cite{Ver21}

%%%%%%%%%%%%%%%%%%%%%%%%
%%% ACKNOWLEDGEMENTS %%%
%%%%%%%%%%%%%%%%%%%%%%%%
\section*{Acknowledgements}
XB and DJ thank the ANR for financial support in the framework of the BSE-forces grant.
PFL thanks the European Research Council (ERC) under the European Union's Horizon 2020 research and innovation programme (grant agreement no.~863481) for financial support. 
DJ is indebted to the CCIPL computational center installed in Nantes for (the always very) generous allocation of computational time. MC and XB acknowledges  HPC resources from GENCI-IDRIS Grant 2020-A0090910016.  
%%%%%%%%%%%%%%%%%
%%% SUPP INFO %%%
%%%%%%%%%%%%%%%%%
%\begin{suppinfo}
\section*{Supporting Information Available}
MO combination and oscillator strengths. Representation of the MOs. Comparison of various CT metrics. Extra details for the dipeptides. Cartesian coordinates for all compounds.
%\end{suppinfo}

%%%%%%%%%%%%%%%%%%%%
%%% BIBLIOGRAPHY %%%
%%%%%%%%%%%%%%%%%%%%
\bibliography{biblio-new}

\end{document}